\documentclass[12pt]{iopart}
\usepackage{graphicx}

\usepackage{iopams}

\expandafter\let\csname equation*\endcsname\relax
\expandafter\let\csname endequation*\endcsname\relax
\usepackage{amsmath}
\usepackage{mathtools}
 
\begin{document}

\title[Evolution of dust ion acoustic soliton in the presence of superthermal electrons]{Evolution of dust ion acoustic soliton in the presence of superthermal electrons}

\author{D. Dutta$^{1,2}$, S. Adhikari$^{2}$, R. Moulick$^{3}$, and K. S. Goswami$^{2}$}

\address{$^1$Physics Department, Gauhati University, Guwahati, Kamrup, Assam-781014, India}
\address{$^2$Centre of Plasma Physics-Institute for Plasma Research, Nazirakhat, Sonapur, Kamrup, Assam-782402, India}
\address{$^3$Lovely Professional University, Jalandhar Delhi- GT Road, Phagwara, Punjab -144411, India}
\ead{dutta3dharitree@gmail.com}
\vspace{10pt}
%\begin{indented}
%\item[]April 2019
%\end{indented}

\begin{abstract}
Propagation of solitary wave in dusty plasmas started to draw the attention of the physicists since the early 90s. The presence of superthermal particles seems to have a great impact on such waves, as they indicate the existence of non-thermal systems. It has been observed that the superthermal population is capable of altering the nature of the plasmas waves. In the present paper, the effect of the superthermal electron population on the dust ion acoustic solitary wave has been explored. The plasma is considered un-magnetized and composed of two components of superthermal electrons (of two distinct temperature) along with positive ions, and negative dust particles. A major part of the work has been concentrated on the stability of the solitary structures considering the effect of the superthermal parameter. In addition, the dust charge has been considered as a variable and a detailed analysis has been provided on the same. The proposed plasma model is most suitable for analyzing Saturn magnetosphere and can be extended to any space plasmas with superthermal population. 
\end{abstract}

%\maketitle

\section{Introduction}

In the early 1960, the space based observations \cite{Vasyliunas,Montgomery,Feldman} reported the presence of superthermal particles in the space and atmospheric environments. Plasma containing  superthermal particles indicates its non-thermal equilibrium state. Such particles are usually found in low-dense ($n~\sim10^6~m^{-3}$), high temperature ($T>10^5~K$)  conditions in space, where, the binary collision of the charged particles is sufficiently rare\cite{Pierrard,Lazar_Schli}. Solar wind is the first, and till now the only stellar outflow that has been measured in-situ, revealing important information about the existence and generation of these superthermal particles. These particles are observed to have a strong influence in wave-particle interactions, and solar events. There are planty of astrophysical phenomena where, superthermal particles (ions and electrons) are found in great abundance providing ample information about their origin. 

The velocity distribution function (VDF) of the superthermal particles is observed to be quasi-Maxwellian within the limit of mean thermal velocity. However, at higher velocities it becomes non-Maxwellian\cite{Pierrard,Lazar_Schli,Maksimovic}. The generalized Lorentzian or kappa distribution function is very much convenient to represent such VDFs as it fits both thermal as well as superthermal part of the velocities. The kappa distribution function is characterized by a parameter $\kappa$, known as the spectral index, which represents the degree of superthermality. The distribution function was formulated by Vasyliunas \cite{Vasyliunas} and is expressed as,

\begin{equation}\label{kappa}
f_s^{\kappa}(v_i) = \frac{n_{s0}}{\left(\pi\kappa \theta_{s}^2\right)^{3/2}}\frac{\Gamma(\kappa+1)}{\Gamma(\kappa-1/2)}\left[1+\frac{v_s^2}{\kappa \theta_{s}^2}\right]^{-(\kappa+1)}.
\end{equation}

Here, $\theta_s$ is the effective thermal velocity of the plasma species $s$ and is expressed by $\theta_{s}^2 = (1-\frac{3}{2}\kappa)v_{T_{s}}^2$, where $v_{T_s} = \sqrt{T_s \mathbin{/} m_s}$. $m_s$ and $v_s$ are the mass and velocity of the particles with characteristic kinetic temperature $T_s$ (the temperature of the equivalent Maxwellian with the same average kinetic energy). $n_{s0}$ is their equilibrium number density. $\Gamma(x)$ is the Gamma function. In the limit $\kappa \rightarrow \infty$, the distribution converges to the Maxwellian distribution. Small value of $\kappa$ indicates the population of highly superthermal particles.

It has been observed that the presence of superthermal particle brings a remarkable change in the nature of the waves in the plasma\cite{Lazar,Alam}. The data from the instruments onboard Cassini (CAPS/ELS ($0.6~eV$ to $26~keV$) and MIMI/LEMMS ($15~keV$ to $10~MeV$)) reported the presence of two components of superthermal electrons with two distinct temperatures in Saturn magnetosphere \cite{Schippers}. In addition, one of the components is found to be less superthermal than the other. The present study aims to investigate the nature of nonlinear structures in the presence of such superthermal electrons and thereby understanding the dynamics of plasma waves in such systems.

Another salient feature of the present work is to incorporate the discrete dust charging model to verify the dust charging condition\cite{Chunshi}. The incorporation of dynamic dust charging is extremely important since the presence of charged dust significantly affects the dispersive properties of ion acoustic waves\cite{Baluku,Moulick,Kakati}. Moreover, the presence of charged dust grains introduces new and different kinds of low frequency wave including dust ion acoustic waves (DIAWs)\cite{Dutta,Mishra1}, dust acoustic waves (DAWs)\cite{Mishra1,Shukla_mamun,Mishra2} etc. Shukla, and Silin’s \cite{Shukla_Silin} work on dust ion acoustic waves in un-magnetized plasma involves highly charged dust particles with the consideration of constant dust charges. However, such assumption becomes ineffective in a realistic situation. Nejoh \cite{Nejoh} pointed out that the dust charge variation affects the characteristics of the collective motion of the plasma particles. In a similar note, Melands\o{}\cite{Melands} also mentioned that a phase difference between the dust charge variation and the wave can lead to strong damping of the wave. Present study checks upon such assumptions where superthermal particles are found.

One of the most common nonlinear structures supported by dusty plasma is soliton. Under suitable environment, dusty plasma can support dust ion acoustic (DIA) and dust acoustic (DA) soliton. A soliton is a special kind of solitary wave which preserves its shape and size even after the interaction. The balance between the nonlinearity and dispersion leads to the formation of these stationary, localized structures. In the absence of dispersion, or if the dispersion fails to compete with the nonlinearity, the amplitude would grow and accelerate, eventually leading to wave breaking. In this case, the peak overtakes the trough, and the wave becomes multiple valued\cite{Swanson}. On other hand, in absence of nonlinearity, the amplitude would gradually decay, eventually leading to wave damping, or the wave gradually disperses away.

The present work intends to explore the existence and nature of solitary waves in dusty plasma containing two components of superthermal electrons along with positively charged cold fluid ion and negatively charged dust.  The standard KdV equation for the system has been derived using the reductive perturbation method \cite{Shukla_book} in order to develop a stationary solution of soliton structure. The stationary solution is considered as the initial solution for the temporal evolution of solitary structure. The effects of superthermal particles on the temporal evolution of different parameters of the soliton are given major importance in the present study. 

This paper is organized as follows. The basic equations and modeling are presented in section \ref{Section_2} along with the detailed analysis of self-consistent dust charge variation, derivation of KdV equation, and linear wave analysis. The numerical methods and parameter details are provided in section \ref{Section_3}. The results and discussions are provided in section \ref{Section_4}. The work is finally concluded in section \ref{Section_5}.

\section{Theoritical formulation} \label{Section_2} 

For nano-sized dust particles, the dust to ion mass ratio\cite{Masood} is typically of the order of $10^3~-~10^4$. Under such situations, the phase velocity of the DIA wave is found to be higher than the dust and ion thermal velocities. Therefore, consideration of the ion and dust particles as cold fluids is quite justified. In addition, the effect of gravitational force acting on the dust particles can be easily neglected due to the size. The plasma under study is unmagnetized and periodic in nature. Hence, the modeling has been performed in one-dimension. The set of governing equations for the system can be written as,

\begin{equation}
\frac{\partial n_i}{\partial t} + \frac{\partial {(n_iv_i)}}{\partial x}  =  0, \label{3 coni}
\end{equation}
\begin{equation}
\frac{\partial v_i}{\partial t} + v_i\frac{\partial v_i}{\partial x}  =  -\frac{e}{m_i}\frac{\partial \phi}{\partial x}, \label{3 momi}
\end{equation}
\begin{equation}
\frac{\partial n_d}{\partial t}+ \frac{\partial {(n_dv_d)}}{\partial x}  =  0, \label{3 cond}\\
\end{equation}
\begin{equation}
\frac{\partial v_d}{\partial t} + v_d\frac{\partial v_d}{\partial x}  = \frac{z_de}{m_d}\frac{\partial \phi}{\partial x}, \label{3 momd}
\end{equation}
\begin{equation}\label{3 poisson}
\frac{\partial ^2 \phi}{\partial x^2}=\frac{e}{\epsilon_0}(n_{ec} +n_{eh} +z_d n_d -n_i).
\end{equation}

The number densities of ion and dust are $n_i$ and $n_d$ with their respective velocities are $v_i$, and $v_d$. $m_i$, and $m_d$ are the masses of ion and dust. $z_d$ represents the number of dust charge. $e$ is the usual electric charge, and $\phi$ is the electric potential.

The two components of the superthermal electron are assumed to follow kappa-distribution. To distinguish these electron components, they have been termed as ‘hot’ and ‘cold’ based on their temperature. The densities of the superthermal electrons can be derived by performing volume integration of the generalized Lorentzian or the kappa distribution function (equation (\ref{kappa})) and formulated as,

\begin{equation}
n_{ec}=n_{0c}\left(1-\frac{e\phi}{T_{ec}(\kappa_c-\frac{3}{2})}\right)^{-\kappa_c+\frac{1}{2}}, \label{3 den cold}\\
\end{equation}
\begin{equation}
n_{eh}=n_{0h}\left(1-\frac{e\phi}{T_{eh}(\kappa_h-\frac{3}{2})}\right)^{-\kappa_h+\frac{1}{2}}. \label{3 den hot}
\end{equation}

Here, $n_{ec}$ and $n_{eh}$ are the densities of the cold and hot components of electron with their equilibrium densities $n_{0c}$, and $n_{0h}$. Their respective spectral indices are symbolized by $\kappa_c$, and $\kappa_h$, and their temperatures are $T_{ec}$, and $T_{eh}$ in energy scale. The electron density has been considered as $\sim~~10^5~~m^{-3}$. The temperatures for the cold and the hot components of electron are taken as $T_{ec}<100~~eV$ and $T_{eh}~~\sim~~100~eV~-~10~keV$. The formulation has been adapted from few earlier works\cite{Pierrard,Lazar_Schli}.

Equations (\ref{3 coni}) - (\ref{3 den hot}) have been normalized with appropriate parameters as below,

$$ \begin{array}{cc}
N_s = \frac{n_s}{n_{i0}},~~~~\mu_s = \frac{n_{s0}}{n_{i0}},~~~~V_s = \frac{v_s}{c_d},~~~~\sigma_s = \frac{T_{ef}}{T_s},\\~\\
Z_d=z_d/z_{d0},~~~~\Phi = \frac{e\phi}{T_{ef}},~~~~\xi = \frac{x}{\lambda_{Dd}},~~~~\tau = \frac{t}{\omega_{pd}^{-1}},
\end{array}$$

where, $s$ indicates individual charged species. Therefore, $N_{ec}$, $N_{eh}$,$N_i$, and $N_d$ are the normalized densities of cold electron, hot electron, ion, and dust. The normalized velocities of those species are respectively $V_{ec}$, $V_{eh}$, $V_i$, and $V_d$. The normalized potential is denoted by $\Phi$. $\zeta$, and $\bar{\tau}$ are the normalized length and time. The velocities are normalized with the dust acoustic velocity ($C_d=\sqrt{T_{ef} \mathbin{/} m_d}$). The system length and time are normalized by dust Debye length ($\lambda_{Dd}=\sqrt{\epsilon_0 T_{ef}\mathbin{/}(z_{d0} n_{d0}e^2)}$) and inverse dust plasma frequency ($\omega_{pd}^{-1}=\sqrt{\epsilon_0m_d \mathbin{/}(n_{d0}z_{d0}^2e^2)}$) respectively. $T_{ef}$ is the effective temperature of the electrons and is given by $T_{ef}=T_{ec}T_{eh} \mathbin{/}(\mu_cT_{eh}+\mu_hT_{ec})$. $z_{d0}$ is the equilibrium dust charge number. After normalization, equations (\ref{3 coni}) - (\ref{3 den hot}) can be written as,

\begin{equation}
\frac{\partial N_i}{\partial \bar{\tau}} + \frac{\partial {(N_iV_i)}}{\partial \zeta}  =  0,  \label{3 coni norm}\\
\end{equation}
\begin{equation}
\frac{\partial V_i}{\partial \bar{\tau}} + V_i\frac{\partial V_i}{\partial \zeta} = -\frac{1}{\delta}\frac{\partial \Phi}{\partial \zeta}, \label{3 momi norm}\\
\end{equation}
\begin{equation}
\frac{\partial N_d}{\partial \bar{\tau}}+ \frac{\partial {(N_dV_d)}}{\partial \zeta}  =  0, \label{3 cond norm}\\
\end{equation}
\begin{equation}
\frac{\partial V_d}{\partial \bar{\tau}} + V_d\frac{\partial V_d}{\partial \zeta}  =  Z_d\frac{\partial \Phi}{\partial \zeta}, \label{3 momd norm}
\end{equation}
\begin{equation}\label{3 poisson norm}
\frac{\partial ^2 \Phi}{\partial \xi^2}=(N_{ec} +N_{eh} +Z_dN_d -N_i),
\end{equation}
\begin{equation}
N_{ec} =  \mu_c\left(1-\frac{\sigma_c\Phi}{(\kappa_c-\frac{3}{2})}\right)^{-\kappa_c+\frac{1}{2}},  \label{3 den cold norm}
\end{equation}
\begin{equation}
N_{eh} = \mu_h\left(1-\frac{\sigma_h\Phi}{(\kappa_h-\frac{3}{2})}\right)^{-\kappa_h+\frac{1}{2}}, \label{3 den hot norm}
\end{equation}

where, $\delta$ is the ion to dust mass ratio ($m_i \mathbin{/}m_d$). $\mu_c$ and $\mu_h$ are the normalized equilibrium electron densities and $\mu=Z_{d0}n_{d0}\mathbin{/}n_{i0}$ is the normalized equilibrium dust density. $\sigma_c$ ($=T_{ef}\mathbin{/}T_{ec}$) and $\sigma_h$ ($T_{ef}\mathbin{/}T_{eh}$) are the temperature ratios of the superthermal electrons. To study the DIA solitons in presence of variable dust charge and no-depleted free electrons \cite{Xie}, $Z_d$ must be reconstructed with the help of the dust charging equations.

\subsection{Self-consistent dust charge variation} %Sub section - 3.2.1

The dynamic dust charging brings a new level of complexity to the system. The charging process is a combination of both electron and ion current; however, there are situations where one might dominate over the other. The charging of dust grains is considered to follow the equation below,

\begin{equation}\label{3 charging}
\frac{dQ_{d}}{dt} = (I_{ec}+I_{eh}+I_{i}).
\end{equation}

Here, $I_{ec}$, $I_{eh}$ and $I_i$ are the cold electron, hot electron, and ion current respectively. In the absence of other effects, due to the higher mobility, electrons will reach the dust grain surface faster than ions and thereby leading to the generation of a negatively charged dust grain population. The dust charging time is of the order of tens of nanosecond while the characteristic time for dust motion is of the order of tens of milliseconds. Thus, on a hydrodynamic time scale, the dust charge can quickly reach local equilibrium. The current balance equation can be written as,

\begin{equation}\label{current balance}
(I_{ec}+I_{eh}+I_{i}) = 0.
\end{equation}

Assuming the streaming velocities of electrons and ions much smaller than their respective thermal velocities, based on OML theory\cite{Shukla_book} one can express the electron and ion current for spherical dust grain with radius $r_d$. Moreover, here the electron species considered are superthermal in nature. Hence, the electron and ion currents\cite{Hakimi,Bora} can be expressed as,
\begin{equation}\label{cold_ele_current}
\begin{multlined}
I_{ec} = -er_{d} ^{2} \mu_c \left(\frac{8\pi T_{ec}}{m_{e}}\right) ^{\frac{1}{2}} \frac{(\kappa_c - \frac{3}{2})^{\frac{1}{2}}}{\kappa_c (\kappa_c -1)} \frac{\Gamma (\kappa_c +1)}{\Gamma (\kappa_c -\frac{1}{2})} 
\\
\times \left[1-\frac{2e\eta}{m_e(\kappa_c-\frac{3}{2})v_{T_{ec}}^{2}}\right]^{-\kappa_c +1}\left[1-\frac{\sigma_c \Phi}{\kappa_c-\frac{3}{2}}\right]^{-\kappa_c+\frac{1}{2}},
\end{multlined}
\end{equation}

\begin{equation}\label{hot_ele_current}
\begin{multlined}
I_{eh} = -er_{d} ^{2} \mu_h \left(\frac{8\pi T_{eh}}{m_{e}}\right) ^{\frac{1}{2}} \frac{(\kappa_h - \frac{3}{2})^{\frac{1}{2}}}{\kappa_h (\kappa_h -1)} \frac{\Gamma (\kappa_h +1)}{\Gamma (\kappa_h -\frac{1}{2})} 
\\
\times \left[1-\frac{2e\eta}{m_e(\kappa_h-\frac{3}{2})v_{T_{eh}}^{2}}\right]^{-\kappa_h +1}\left[1-\frac{\sigma_h \Phi}{\kappa_h-\frac{3}{2}}\right]^{-\kappa_h+\frac{1}{2}},
\end{multlined}
\end{equation}
\begin{equation}\label{ion_current}
I_{i}=er_{d}^{2}\left(\frac{8\pi T_{i}}{m_i}\right)^{\frac{1}{2}}N_i\left[1-\frac{e\eta}{T_i}\right].
\end{equation}

\begin{figure}
		\centering    
		\includegraphics[width=0.44\textwidth]{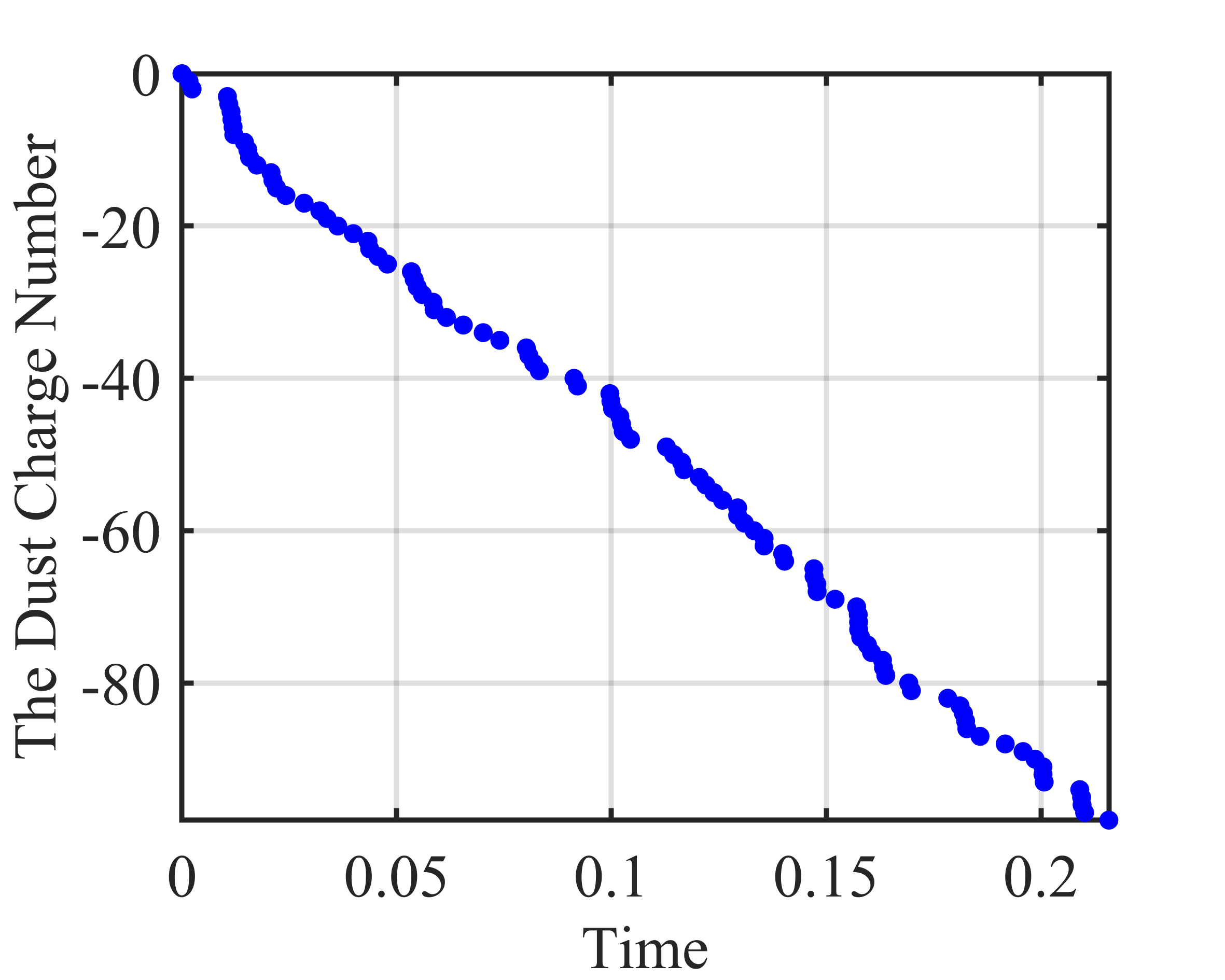}
		\includegraphics[width=0.4\textwidth]{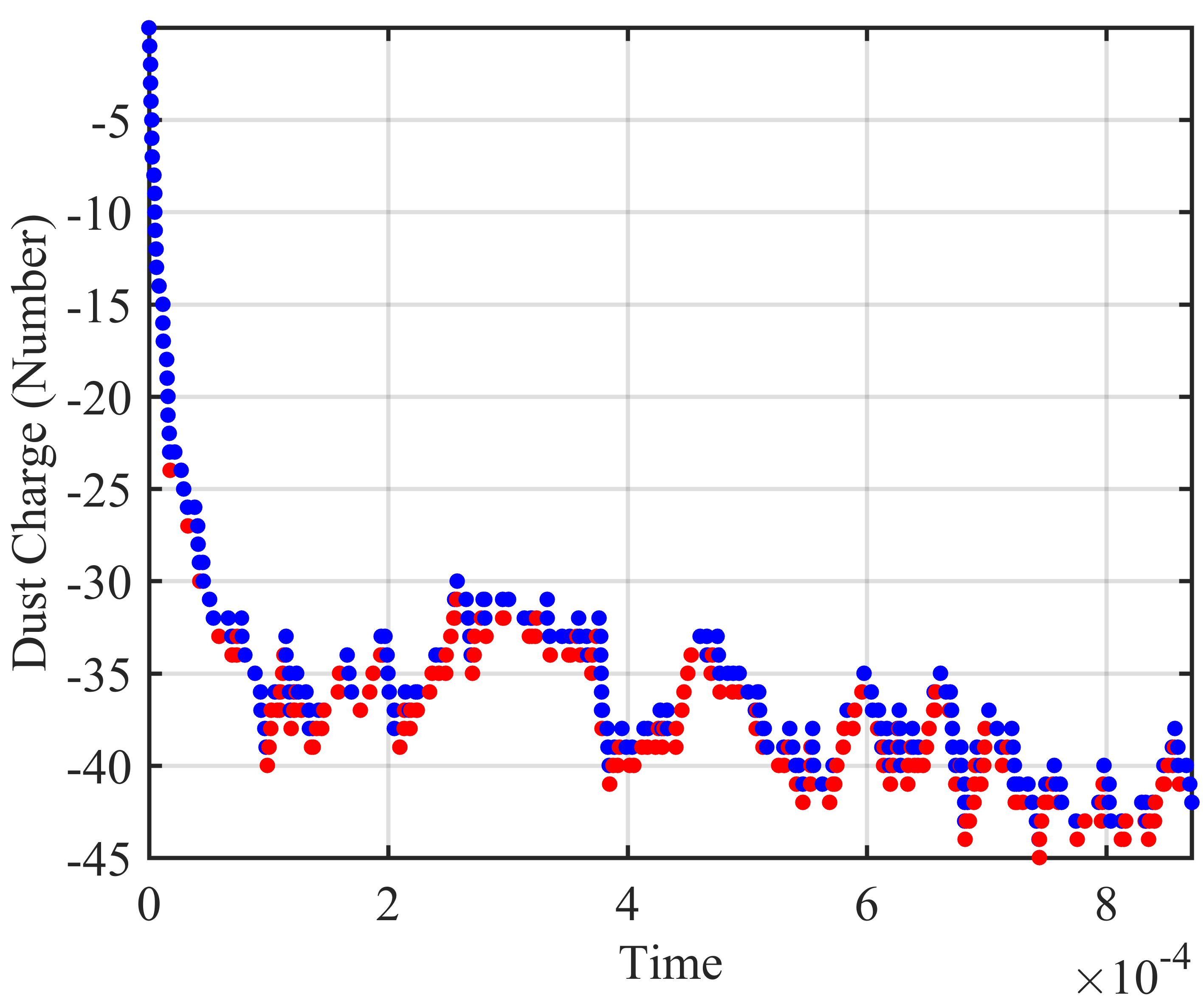}
		\caption[Temporal evolution of the soliton potential for $(a)$ electron ion plasmas, and $(b)$ $\kappa_c=7.0$, $\kappa_h=10.0$, $T_{ec}=10~eV$, $T_{eh}=100~eV$, $\mu_c=0.5$, $\mu_h=0.493$, $\mu=0.007$, and $\delta=0.001$]{The charging of the dust grains using discrete charging model (LEFT) in presence of superthermal electrons ($\kappa_c=7.0$, $T_{ec}=10~eV$, and $\mu_c=0.5$), (RIGHT) in presence of Maxwellian electrons. The blue and red dots represent electrons and ions respectively.}
		\label{fig:dust_charge}
	\end{figure}

Here, $\eta$ is the dust grain surface potential relative to the plasma potential $\Phi$. $v_{T_{ec}}$ and $v_{T_{eh}}$  are the thermal velocities of the corresponding electron components. The other charging processes such as, secondary emission, photoemission, are considered to have negligible contribution in  dust grain charging. Moreover, in the  present scenario, the ion current is extremely small in comparison to the electron current\cite{Duha}, i.e., $I_i\ll I_{ec},I_{eh}$. The assumption has been verified using the discreet charging model of dust developed by Chunsi et al. \cite{Chunshi}. It has been depicted in figure \ref{fig:dust_charge}. The figure in the LHS shows the dominance of electron current for the charging of dust over a time of 0.21 seconds following discrete charging model in the presence of superthermal electrons ($\kappa_c=7$, $T_{ec}=10~eV$, $\mu_c=0.5$). The figure in the RHS represents the charging of the dust grain inpresence of Maxwellian electrons. The blue and red dots represent electrons and ions respectively. Hence, it is reasonable to drop the ion current contribution from the current balance equation (\ref{current balance}). The equation then becomes,
	
	\begin{equation}\label{total current}
	I_{ec} +I_{eh} = 0.
	\end{equation}
	
	Finally, from equations (\ref{cold_ele_current})-(\ref{total current}),
	
	\begin{equation}\label{total current 2}
	\begin{multlined}
	\mu_c\sigma_c^{-\frac{1}{2}}\frac{(\kappa_c -\frac{3}{2})^{\frac{1}{2}}}{\kappa_c(\kappa_c-1)} \frac{\Gamma (\kappa_c+1)}{\Gamma (\kappa_c-\frac{1}{2})}\left(1-\frac{\sigma_c\psi}{\kappa_c-\frac{3}{2}}\right)^{-\kappa_c+1}\left(1-\frac{\sigma_c\Phi}{\kappa_c-\frac{3}{2}}\right)^{-\kappa_c+\frac{1}{2}}\\~\\
	+\mu_h\sigma_h^{-\frac{1}{2}}\frac{(\kappa_h -\frac{3}{2})^{\frac{1}{2}}}{\kappa_h(\kappa_h-1)} \frac{\Gamma (\kappa_h+1)}{\Gamma (\kappa_h-\frac{1}{2})}\left(1-\frac{\sigma_h\psi}{\kappa_h-\frac{3}{2}}\right)^{-\kappa_h+1}\left(1-\frac{\sigma_h\Phi}{\kappa_h-\frac{3}{2}}\right)^{-\kappa_h+\frac{1}{2}}=0
	\end{multlined} 
	\end{equation}
	
	where $\psi=e\eta \mathbin{/} T_{ef}$. Equation (\ref{total current 2}) is important in determining the dust charge $Q_d$, which relates to the dust grain surface potential by $\eta=Q_d\mathbin{/}C$. The capacitance of the spherical dust grain in a plasma can be expressed as, $C=r_d\exp{(-r_d \mathbin{/} \lambda_D)}\approx r_d$ for $\lambda_D\gg r_d$. Hence, the normalized dust charge becomes, $Z_d=\psi\mathbin{/}\psi_0$ , where, $\psi_0=\psi$ ($\Phi=0$) is the dust charge floating potential with respect to the unperturbed plasma potential. $\psi_0$ can be determined by solving the following polynomial,
	
	\begin{equation}\label{psi 0}
	\begin{multlined}
	\left[\mu_c\vartheta_1\sigma_c^2\frac{\kappa_c(\kappa_c-1)}{2(\kappa_c-3/2)^{2}}+\mu_h\vartheta_2\sigma_h^2\frac{\kappa_h(\kappa_h-1)}{2(\kappa_h-3/2)^{2}}\right]\psi_0^{2}\\~\\
	+\left[\mu_c\vartheta_1\sigma_c(\frac{\kappa_c-1}{\kappa_c-3/2})+\mu_h\vartheta_2\sigma_h(\frac{\kappa_h-1}{\kappa_h-3/2})\right]\psi_0+\mu_c\vartheta_1+\mu_h\vartheta_2 = 0,
	\end{multlined} 
	\end{equation}
	
	where, $\vartheta_1 = \sigma_c^{-1/2}\frac{(\kappa_c-3/2)^{1/2}}{\kappa_c(\kappa_c-1)}\frac{\Gamma(\kappa_c+1)}{\Gamma(\kappa_c-1/2)}$, and $\vartheta_2 = \sigma_h^{-1/2}\frac{(\kappa_h-3/2)^{1/2}}{\kappa_h(\kappa_h-1)}\frac{\Gamma(\kappa_h+1)}{\Gamma(\kappa_h-1/2)}$.\\
	
	$\psi_0$ will have two roots with positive and negative value. However, for negatively charged dust, it is always negative. Hence, only the negative values have been taken under consideration. The expression for variable dust charge, $Z_d$ now can be implemented to derive KdV equation for the system.
	
	\subsection{Derivation of the KdV equation} %Sub section - 3.2.2
	
	To study the dynamics of the finite amplitude DIA wave, the KdV equation can be derived from equations (\ref{3 coni norm}) - (\ref{3 den hot norm}) by employing the reductive perturbation technique\cite{Shukla_book}. The stretched coordinate used here are, $\xi=\epsilon^{1/2}(\zeta-M\bar{\tau})$ and $\tau = \epsilon^{3/2}\bar{\tau}$, where $\epsilon$ is the smallness parameter measuring the weakness of the amplitude and $M$ is the speed of the solitary structure. The variables $N_s$, $V_s$, $\Phi$, and $Z_d$ can be expanded about the unperturbed states in the power series of $\epsilon$ as,
	\begin{equation}\label{expansion}
%	\begin{multlined}
	\begin{aligned}
	N_s=N_s^{(0)}+\epsilon N_s^{(1)}+\epsilon^{2}N_s^{(2)}+\epsilon^{3}N_s^{(3)}+\ldots~\ldots\\
	V_s=V_s^{(0)}+\epsilon V_s^{(1)}+\epsilon^{2}V_s^{(2)}+\epsilon^{3}V_s^{(3)}+\ldots~\ldots\\
	\Phi=\epsilon \Phi^{(1)}+\epsilon^{2}\Phi^{(2)}+\epsilon^{3}\Phi^{(3)}+\ldots~\ldots\\
	Z_d=1+\epsilon Z_d^{(1)}+\epsilon^{2}Z_d^{(2)}+\epsilon^{3}Z_d^{(3)}+\ldots~\ldots
%	\end{multlined}
	\end{aligned}  
	\end{equation}
	%	\begin{eqnarray}\label{expansion}
	%	N_s=N_s^{(0)}+\epsilon N_s^{(1)}+\epsilon^{2}N_s^{(2)}+\epsilon^{3}N_s^{(3)}+\ldots~\ldots\\
	%	V_s=V_s^{(0)}+\epsilon V_s^{(1)}+\epsilon^{2}V_s^{(2)}+\epsilon^{3}V_s^{(3)}+\ldots~\ldots\\
	%	\Phi=\epsilon \Phi^{(1)}+\epsilon^{2}\Phi^{(2)}+\epsilon^{3}\Phi^{(3)}+\ldots~\ldots\\
	%	Z_d=1+\epsilon Z_d^{(1)}+\epsilon^{2}Z_d^{(2)}+\epsilon^{3}Z_d^{(3)}+\ldots~\ldots
	%	\end{eqnarray}
	
	Here, ‘$s$’ is used to indicate individual charged species. Therefore, $N_s^{(0)}=1$, for ion and $N_s^{(0)}=\mu$, for dust. After transformation and taking a first order approximation, the equations (\ref{3 coni norm}) - (\ref{3 poisson norm}) become,
	\begin{equation}\label{reductive first order}
	\begin{aligned}
	N_{i}^{(1)} = \frac{1}{\delta M^2}\Phi^{(1)},~~~~V_{i}^{(1)} = \frac{1}{\delta M}\Phi^{(1)},~~~~N_{d}^{(1)} = -\frac{\mu}{M^2}\Phi^{(1)},\\
	V_{d}^{(1)} = -\frac{1}{M}\Phi^{(1)},~~~~ Z_{d}^{(1)} = \gamma_1\Phi^{(1)},\\
	M = \left(\frac{\mu_c+1/\delta}{P_1+Q_1+\mu\gamma_1}\right)^{1/2}
	\end{aligned} 
	\end{equation}
	
	%	\begin{eqnarray}{c}\label{reductive first order}
	%	N_{i}^{(1)} = \frac{1}{\delta M^2}\eta^{(1)},~~~~V_{i}^{(1)} = \frac{1}{\delta M}\eta^{(1)},~~~~N_{d}^{(1)} = -\frac{\mu}{M^2}\eta{(1)},\\
	%	V_{d}^{(1)} = -\frac{1}{M}\eta^{(1)},~~~~ Z_{d}^{(1)} = \gamma_1\Phi^{(1)},\\
	%	M = \left(\frac{\mu_c+1/\delta}{P_1+Q_1+\mu\gamma_1}\right)^{1/2}
	%	\end{eqnarray}
	%\mathbin{/}
	$$
%	\begin{aligned}
	\text{Here, } P_1 = {\mu_c\sigma_c(2\kappa_c-1)}\mathbin{/}{(2\kappa_c-3)}, Q_1 = {\mu_h\sigma_h(2\kappa_h-1)}\mathbin{/}{(2\kappa_h-3)}, \text{and }$$ 
	$$\gamma_1 = -{(\alpha_2+\iota \psi_0)}\mathbin{/}{(\alpha_1 \psi_0+2\beta_1\psi_0)}, \text{with } $$
	$$\iota = \mu_c\vartheta_1\sigma_c^2\frac{(\kappa_c-1)(\kappa_c-1/2)}{2(\kappa_c-3/2)^2}+\mu_h\vartheta_2\sigma_h^2\frac{(\kappa_h-1)(\kappa_h-1/2)}{2(\kappa_h-3/2)^2},$$
	$$\alpha_1 = \mu_c\vartheta_1\sigma_c\left(\frac{\kappa_c-1}{\kappa_c-3/2}\right)+\mu_h\vartheta_2\sigma_h\left(\frac{\kappa_h-1}{\kappa_h-3/2}\right),
	\alpha_2 = P_1\vartheta_1+Q_1\vartheta_2, \text{and }$$ 
	$$\beta_1 =\mu_c\vartheta_1\sigma_c^2\frac{\kappa_c(\kappa_c-1/2)}{2(\kappa_c-3/2)^2}
	+\mu_h\vartheta_2\sigma_h^2\frac{\kappa_h(\kappa_h-1/2)}{2(\kappa_h-3/2)^2}.
%	\end{aligned}
	$$

	The expression for $M$ represents the linear dispersion relation for DIA waves.  The presence of the fluctuating dust charge significantly modifies the relation. To derive the KdV equation for the given system, the expressions with the next higher order in $\epsilon$ are taken into account,
	
	\begin{equation}\label{reductive second order 1}
	\frac{\partial N_i^{(1)}}{\partial\tau}-M\frac{\partial N_i^{(2)}}{\partial\xi}+\frac{\partial\left(N_i^{(1)}V_i^{(1)}\right)}{\partial\xi}+\frac{\partial V_i^{(2)}}{\partial\xi} = 0,
	\end{equation}
	\begin{equation}
	\frac{\partial V_i^{(1)}}{\partial\tau}-M\frac{\partial V_i^{(2)}}{\partial\xi}+V_i^{(1)}\frac{\partial V_i^{(1)}}{\partial\xi} = -\frac{1}{\delta}\frac{\partial\Phi^{(2)}}{\partial\xi},
	\end{equation}
	\begin{equation}
	\frac{\partial N_d^{(1)}}{\partial\tau}-M\frac{\partial N_d^{(2)}}{\partial\xi}+\frac{\partial\left(N_d^{(1)}V_d^{(1)}\right)}{\partial\xi}+\mu\frac{\partial V_d^{(2)}}{\partial\xi} = 0,
	\end{equation}
	\begin{equation}
	\frac{\partial V_d^{(1)}}{\partial\tau}-M\frac{\partial V_d^{(2)}}{\partial\xi}+V_d^{(1)}\frac{\partial V_d^{(1)}}{\partial\xi} = \frac{\partial\Phi^{(2)}}{\partial\xi}+Z_d^{(1)}\frac{\partial\Phi^{(1)}}{\partial\xi},
	\end{equation}
	\begin{equation}
	\frac{\partial^{2}\Phi^{(1)}}{\partial\xi^2} = (P_1+Q_1)\Phi^{(2)}+(P_2+Q_2)\left[\Phi^{(1)}\right]^{2}+N_d^{(2)}+Z_d^{(1)}N_d^{(1)}+\mu Z_d^{(2)}-N_i^{(2)},
	\end{equation}
	\begin{equation}\label{reductive second order 6}
	Z_d^{(2)} = \gamma_1\Phi^{(2)}+\gamma_2\left[\Phi^{(1)}\right]^{2}.
	\end{equation}
	
	$$\text{Here, } P_2 = {\mu_c\sigma_c^2(4\kappa_c^2-1)}\mathbin{/}{2(2\kappa_c-3)^{2}}, Q_2 = {\mu_h\sigma_h^2(4\kappa_h^2-1)}\mathbin{/}{2(2\kappa_h-3)^{2}}, \text{and }$$ 
	$$\gamma_2 = -{(\beta_2+\iota \gamma_1\psi_0+\beta_1\gamma_1^2\psi_0^2)}\mathbin{/}{(\alpha_1\psi_0+2\beta_1\psi_0^2)} \text{ with } \beta_2 = P_2\vartheta_1+Q_2\vartheta_2.$$
	
	By using equations (\ref{reductive second order 1}) - (\ref{reductive second order 6}) and eliminating all the second-order terms, the standard KdV equation is derived,
	
	\begin{equation}\label{KdV}
	\frac{\partial\Psi}{\partial\tau}+a\Psi\frac{\partial\Psi}{\partial\xi}+b\frac{\partial^{3}\Psi}{\partial\xi^{3}}=0,
	\end{equation}
	
	where,$\Psi \equiv \Phi^{(1)}$, and $a$, $b$ represent the nonlinear coefficient, and the dispersion coefficient respectively and are expressed as,
	
	\begin{equation}\label{3 coefficients}
	\begin{multlined}
	a=\left[2P_2+2Q_2-\frac{3}{\delta ^{2}M^3}-\frac{\mu \gamma_1}{M^{2}}-\frac{2\gamma_1}{M^2}+2\gamma_2\mu\right]\left[-\frac{2\mu}{M^2}-\frac{2}{\delta M^{3}}\right]^{-1},\\~\\ 
	b = \left[\frac{2\mu}{M^3}+\frac{2}{\delta M^3}\right]^{-1}.
	\end{multlined} 
	\end{equation}
	%	\begin{eqnarray}\label{3 coefficients}
	%	$a=[2P_2+2Q_2-\frac{3}{\delta ^{2}M^3}-\frac{\mu \gamma_1}{M^{2}}-\frac{2\gamma_1}{M^2}+2\gamma_2\mu][-\frac{2\mu}{M^2}-\frac{2}{\delta M^{3}}]^{-1}$,\\~\\ 
	%	$b = [\frac{2\mu}{M^3}+\frac{2}{\delta M^3}]^{-1}$.
	%	\end{eqnarray}
	
%	Equation \ref{KdV} can be written as
%	
%	\begin{equation}\label{KdV final}
%	\frac{\partial\Phi}{\partial\tau}+a\Phi\frac{\partial\Phi}{\partial\xi}+b\frac{\partial^{3}\Phi}{\partial\xi^{3}}=0
%	\end{equation}
	
	The stationary solution of equation (\ref{KdV}) is
	
	\begin{equation}\label{3 soliton}
	\Psi = \chi \text{ sech}^2[(\xi - U\tau)\mathbin{/}\Delta], 
	\end{equation}
	where, $\chi = 3U/a$, $\Delta = 2\sqrt{b/U}$ are the amplitude and width of the soliton and $U$ is the velocity. Equation (\ref{3 soliton}) clearly indicates that the plasma can support any soliton only when $b>0$ for $U>0$. It is obvious from equations (\ref{3 coefficients}) - (\ref{3 soliton}) that the DIA wave may support rarefactive and (or) compressive soliton depending upon the value of $a$. For $a>0$, it can support soliton with positive polarity and vice versa. The variation of the nonlinearity and dispersion coefficient with the cold electron concentration for different spectral indices are depicted in figure \ref{fig:coefficients}.
	
	Following the same procedures, the KdV equations for the ion and dust number density are derived. By dropping the power $(1)$ from the equations for simplicity\cite{Kakati} (i.e., assuming $N_{i}^{(1)} \equiv N_{i}$ and $N_{d}^{(1)} \equiv N_{d}$), the KdV euations can be expressed as,
	
	\begin{equation}\label{KdVi}
	\frac{\partial N_i}{\partial\tau}+cN_i\frac{\partial N_i}{\partial\xi}+d\frac{\partial^3N_i}{\partial\xi^3} = 0,
	\end{equation}
	\begin{equation}\label{KdVd}
	\frac{\partial N_d}{\partial\tau}+fN_d\frac{\partial N_d}{\partial\xi}+g\frac{\partial^3N_d}{\partial\xi^3} = 0.
	\end{equation}
	
	Here, $$c=\frac{3M-2\delta^2M^5(P_2+Q_2)-\frac{1}{\mu}(3M-3\gamma_1M^3+2\gamma_2M^5)\left\{\frac{1-\delta M^2(P_1+Q_1)}{1-\gamma_1M^2}\right\}^2}{2\left\{1-\left(\frac{1-\delta M^2(P_1+Q_1)}{1-\gamma_1M^2}\right)\right\}},$$
	$$d=g=\frac{\delta M^3}{2\left\{1-\left(\frac{1-\delta M^2(P_1+Q_1)}{1-\gamma_1M^2}\right)\right\}}\text{, and}$$  $$f=c\times\left\{\frac{1-\gamma_1M^2}{1-\delta M^2(P_1+Q_1)}\right\}.$$
	
	The solution of equation (\ref{KdVi}) and (\ref{KdVd}) takes the following form,
	
	\begin{equation}\label{solitoni}
	N_i=(3U/c)\text{ sech}^2\left[(\xi-U\tau)\sqrt{U/2d}\right],
	\end{equation}
	\begin{equation}\label{solitond}
	N_d=(3U/f)\text{ sech}^2\left[(\xi-U\tau)\sqrt{U/2g}\right].
	\end{equation}
	
	\subsection{The linear wave analysis} %Sub section - 3.2.3
	
	The general dispersion relation for the electrostatic wave in a plasma consisting of kappa-distributed species\cite{Baluku} can be written as,
	
	\begin{equation}\label{dispersion 3 1}
	D(k,\omega) = 1-\sum_{ec,eh,i,d}\frac{\omega_{ps}^2}{k^2\theta_s^2}\mathcal{Z}'(\kappa_s; \varsigma_s) = 0,
	\end{equation}
	
	where, $s$ represents the cold ($ec$) and hot ($eh$) components of electron, ions ($i$), and dust particles ($d$) respectively. $\varsigma_s=\omega/(k\theta_s)$ (for superthermal particles), and $\omega/(kv_s)$ (for Maxwellian particles), are the normalized complex wave phase velocities of species $s$. Here, $\omega=\omega_r+i\gamma$ is the complex angular frequency, and $k$ is the wave number. $\mathcal{Z}'(\kappa_s; \varsigma_s)$ is the derivative\cite{Lazar,Baluku} of the modified plasma dispersion function $\mathcal{Z}(\kappa_s; \varsigma_s)$ with respect to $\varsigma_s$, and is expressed as
	\begin{equation}\label{dispersion 3 2}
	\mathcal{Z}'(k_s; \varsigma_s)= -\frac{(\kappa_s-1/2)}{\kappa_s(\kappa_s+1)}{}_2F_1\left[2,2\kappa_s+1,\kappa_s+2;\frac{1}{2}(1+i\varsigma_s/\sqrt{\kappa_s})\right].
	\end{equation}
	
	Here, ${}_2F_{1}$ is the Gauss Hypergeometric function\cite{Abramowitz}. In the limit, $\kappa_s\rightarrow\infty$, $\mathcal{Z}(\kappa_s;\varsigma_s)$ reduces to the usual plasma dispersion relation $Z(\varsigma)$ introduced by Fried and Conte\cite{Fried}. On the dust ion acoustic time scale, $\varsigma_d,\varsigma_i\gg1$, and $\varsigma_{ec, eh}<1$. Hence, using the asymptotic expansion for dusts and ions, and the power series expansion for electrons, the dispersion relation can be written as,
	\begin{equation}\label{dispersion 3 3}
	\begin{multlined}
	1+\sum_{j=ec, eh} \frac{\mu_j\sigma_j}{k^{2}\lambda_{D}^{2}}\left(\frac{\kappa_j}{\kappa_j-3/2}\right)\left[\frac{2i\sqrt{\pi}}{\sqrt{\kappa_j}}\frac{\omega_{r}}{k\theta_j}+\frac{(2\kappa_j-1)}{\kappa_j}\right]\\
	-\frac{\omega_{p_{i}}^2}{\omega_r^2}\left[\left\{1+\frac{3k^2\theta_i^2}{\omega_r^2}\left(\frac{\kappa_i}{2\kappa_i-3}\right)\right\}+\frac{2i\sqrt{\pi}}{\sqrt{\kappa_i}}\frac{\Gamma(\kappa_i)}{\Gamma(\kappa_i-1/2)}\frac{\omega_r^2}{k^2\theta_i^2}\left(1+\frac{1}{\kappa_i}\frac{\omega_r^2}{k^2\theta_i^2}\right)^{-(\kappa_i+1)}\right]\\
	-\frac{k^2\theta_d^2}{\omega_r^2} = 0 
	~~~~~~\mbox{for} ~~~~~~ \text{Im}(\varsigma_s)>0.
	\end{multlined} 
	\end{equation}
	%	\begin{equation}\label{dispersion 3 3}
	%	$$1+\sum_{j=ec, eh} \frac{\mu_j\sigma_j}{k^{2}\lambda_{D}^{2}}\frac{\kappa_j}{\kappa_j-3/2}[\frac{2i\sqrt{\pi}}{\sqrt{\kappa_j}}\frac{\omega_{r}}{k\theta_j}+\frac{(2\kappa_j-1)}{\kappa_j}]-\frac{\omega_{p_{i}}^2}{\omega_r^2}[\{1+\frac{3k^2\theta_i^2}{\omega_r^2}(\frac{\kappa_i}{2\kappa_i-3})\}+$$
	%	\vspace{0.2cm}
	%	$$ \frac{2i\sqrt{\pi}}{\sqrt{\kappa_i}}\frac{\Gamma(\kappa_i)}{\Gamma(\kappa_i-1/2)}\frac{\omega_r^2}{k^2\theta_i^2}(1+\frac{1}{\kappa_i}\frac{\omega_r^2}{k^2\theta_i^2})^{-(\kappa_i+1)}]-\frac{k^2\theta_d^2}{\omega_r^2} = 0 $$
	%	\end{equation}
	
	Here, $\lambda_D = \sqrt{ \epsilon_0 T_{ef}/n_{i0} e^2}$ is the effective Debye length and $\omega_{pi}= \sqrt{n_{i0} e^2/ \epsilon_0 m_i}$ is the ion plasma frequency. The spectral index and effective thermal velocity becomes $\kappa_i \rightarrow \infty$, and $\theta_i\rightarrow{T_{i}}$ respectively. Separating the real and imaginary part of the equation (\ref{dispersion 3 3}), the dispersion relation of the given system as well as the growth/ damping of the DIA wave can be determined. Therefore, the real part of equation (\ref{dispersion 3 3}) becomes,
\begin{equation}\label{dispersion 3 4}
\begin{multlined}
1+\left[\mu_c\sigma_c\left(\frac{\kappa_c-1/2}{\kappa_c-3/2}\right)+\mu_h\sigma_h\left(\frac{\kappa_h-1/2}{\kappa_h-3/2}\right)\right]\frac{1}{k^2\lambda_D^2}\\
-\frac{1}{\left(\omega_r/\omega_{pi}\right)^2}\left[1+\frac{3k^2\lambda_D^2}{\left(\omega_r/\omega_{pi}\right)^2}\right]
-\frac{\mu\delta}{\left(\omega_r/\omega_{pi}\right)^2}=0.
\end{multlined}
\end{equation}

For weakly damped or amplified wave, assuming $|\gamma|\ll\omega_r$ the damping or growth rate can be expressed as,
\begin{equation}\label{dispersion 3 5}
\begin{multlined}
\frac{|\gamma|}{\omega_{pi}}=\sqrt{\frac{\pi}{8}}\sqrt{\frac{m_e}{m_i}}\left[\mu_c\sigma_c^{3/2}\frac{\Gamma(\kappa_c+1)}{\Gamma(\kappa_c-1/2)(\kappa_c-3/2)^{3/2}}+\mu_h\sigma_h^{3/2}\frac{\Gamma(\kappa_h+1)}{\Gamma(\kappa_h-1/2)(\kappa_h-3/2)^{3/2}}\right]\\
\frac{\left(\omega_r^4/\omega_{pi}^4\right)}{k^3\lambda_D^3}+
\sqrt{\frac{\pi}{8}}\sigma_i^{3/2}\frac{\Gamma(\kappa_i+1)}{\Gamma(\kappa_i-1/2)(\kappa_i-3/2)^{3/2}}\left[1+\frac{\sigma_i}{2(\kappa_i-3/2)}\frac{\left(\omega_r^2/\omega_{pi}^2\right)}{k^2\lambda_D^2}\right]^{-(\kappa_i+1)}.
\end{multlined} 
\end{equation}

Here, $\sigma_i=T_{ef}/T_i$. The growth or damping rate can be estimated by plotting the dispersion curve $(k\lambda_D - |\gamma|/\omega_{pi})$. The solution of the imaginary part of the dispersion relation (equation (\ref{dispersion 3 3})) will be helpful to understand whether the wave is amplified or damped. A positive $\gamma$ indicates an exponentially growing wave whereas a negative $\gamma$ stands for the damping\cite{Chen}.

%********************************** %Third Section  **************************************
\section{Numerical method and parameter details} \label{Section_3}

The evolution of the solitary wave is observed by numerically solving the equation (\ref{KdV}) using Fourier Spectral method \cite{Kassam} with an initial solution, $(3U/a)\text{ sech}^2\left[(\xi-U\tau)\sqrt{U/2b}\right]$ (equation (\ref{3 soliton})). The approach has been adapted from the  work of Mishra et al.\cite{Mishra_Corr}, which has been found to be an efficient way to evaluate solitary wave propagation. The time interval has been taken as $d\tau=5\times10^{-4}$. The boundary is considered periodic. For potential the following parameters have been used, $\kappa_c=7$, $\kappa_h=10$, $T_{ec}=10~eV$, $T_{eh}=100~eV$, $\delta=10^{-3}$, $\mu=0.007$, $\mu_c=0.5$, and $\mu_h=0.493$. For the evolution of the dust density the parameters have been kept alike. However, for dust density, the parameters are also varied separately. Details have been provided in their respective sections (section \ref{Section_4}).

%********************************** %Fourth Section  **************************************
\section{Results and discussions} \label{Section_4}

\begin{figure}
	\centering    
	\includegraphics[width=0.6\textwidth]{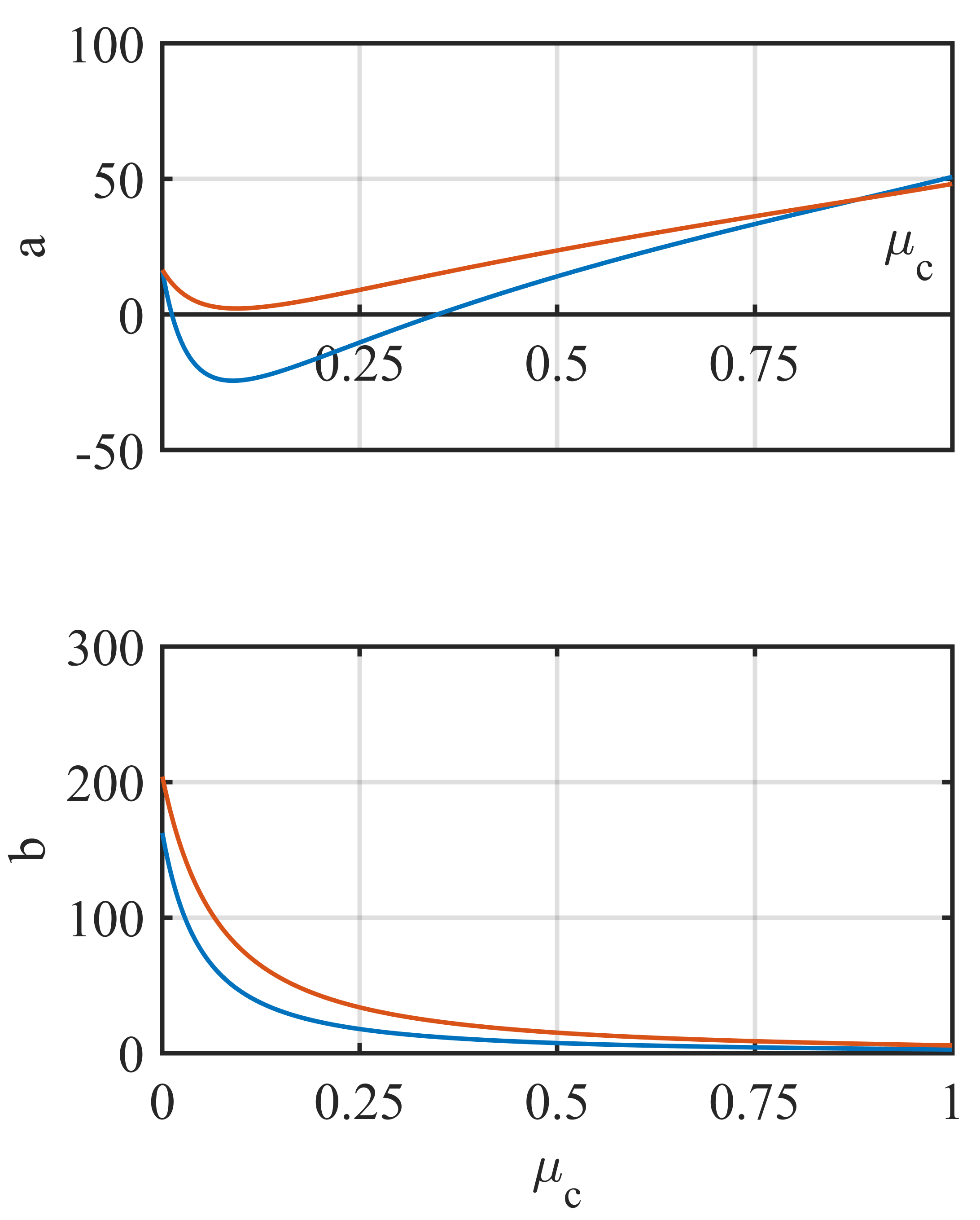}
	\caption[The nonlinear coefficient $a$ (upper panel) and the dispersion coefficient $b$ (lower panel) of the KdV equation. The curve coding in both of the panels: $(a)$ $\kappa_c=7.0$, $\kappa_h=10.0$ (blue line), $(b)$ $\kappa_c=100.0$, $\kappa_h=100.0$ (red line)]{The nonlinear coefficient $a$ (upper panel) and the dispersion coefficient $b$ (lower panel) of the KdV equation [equation (\ref{KdV})]. The curve coding in both of the panels: $(a)$ $\kappa_c=7.0$, $\kappa_h=10.0$ (blue line), $(b)$ $\kappa_c=100.0$, $\kappa_h=100.0$ (red line).}
	\label{fig:coefficients}
\end{figure}

Solitons are the result of the balance between the forces that control the dynamical evolution of a nonlinear system. However, the presence of superthermal particle may alter the dynamic behavior of such nonlinear system. To investigate the effect of superthermal parameters, the nonlinear and dispersion coefficients of the governing equation have been evaluated. In figure \ref{fig:coefficients}, the top figure depicts the behavior of the nonlinear coefficient while the bottom represents the behavior of the dispersion coefficient. It has been observed that the plasma can support both compressive and rarefactive soliton depending upon the spectral indices and the concentration of the superthermal species. For superthermal electrons ($\kappa_c=7.0$, $\kappa_h=10.0$), there is a range of densities ($\mu_c=0.01~-~0.35$) within which the solitons will be rarefactive and beyond will be compressive. Only compressive solitons are observed in plasmas containing non-superthermal electrons ($\kappa_c=100.0$, $\kappa_h=100.0$).

From equation (\ref{KdV}), it is clear that the amplitude of the soliton is inversely proportional to the nonlinear coefficient. The presence of non-superthermal electrons in the plasma initially reduces the amplitude of the soliton for lower values of $\mu_c$, and then raises it up for higher values. However, for superthermal electrons, the variation in amplitude with the values of $\mu_c$ is different for the different modes of polarity (solid line, figure \ref{fig:coefficients}). With the increase of the spectral indices the amplitude of the soliton would increase, regardless the densities of different electron population.

The figure at the bottom panel represents the variation of dispersion coefficient with cold electron concentration for the aforesaid two sets of spectral indices. Since, the dispersion coefficient is directly proportional to the width of the soliton, one can understand the variation of the width of the soliton with different values of $\kappa_c$, $\kappa_h$, and $\mu_c$. On the increase of the spectral indices, the width of the soliton increases, however, it decreases with an increase in $\mu_c$. It is to be mentioned that the temperature for the cold and hot components of electron are considered as $T_{ec}=10~eV$, and $T_{eh}=100~eV$.

\begin{figure}
	\centering    
	\includegraphics[width=0.8\textwidth]{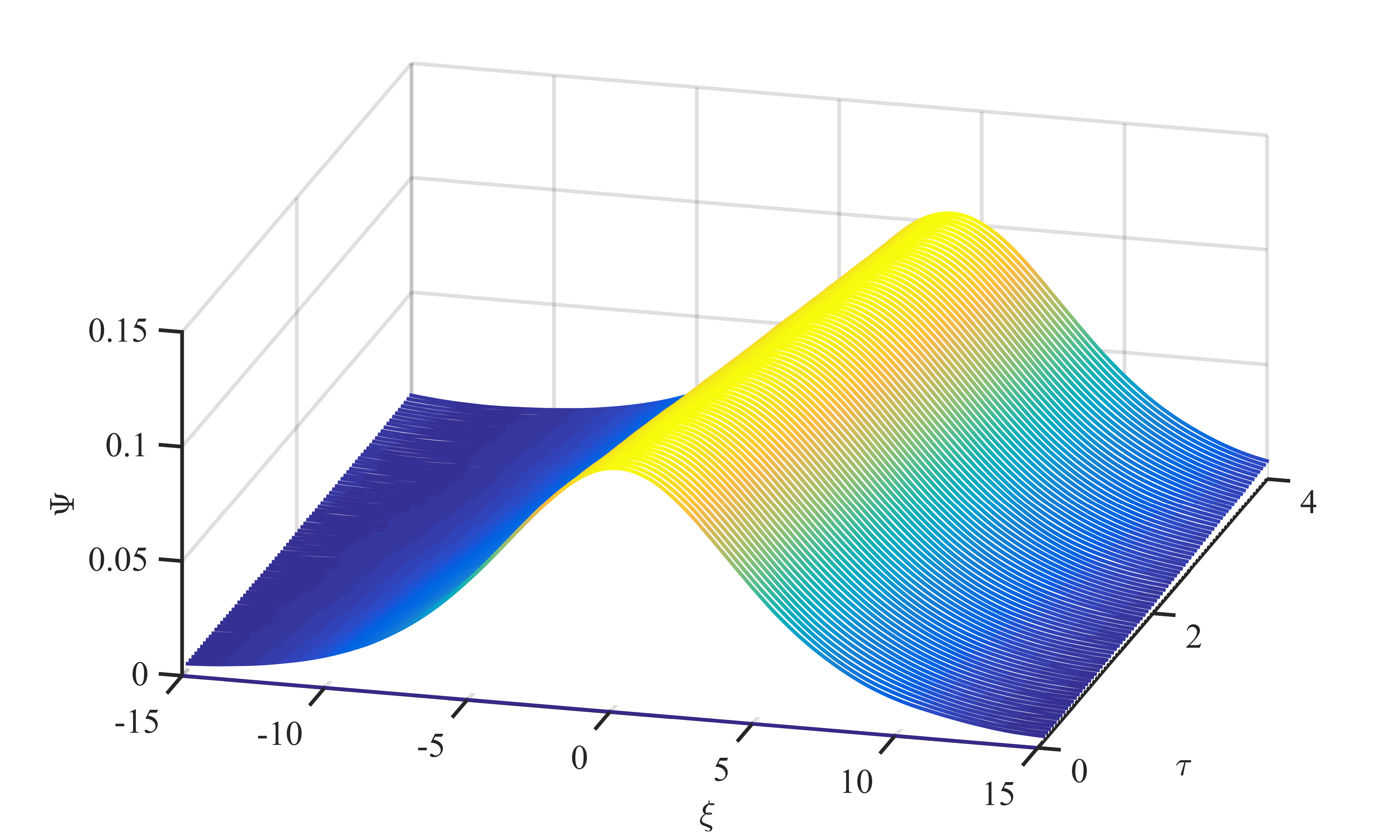}
	\caption[Temporal evolution of the soliton potential for $\kappa_c=7.0$, $\kappa_h=10.0$, $T_{ec}=10~eV$, $T_{eh}=100~eV$, $\mu_c=0.5$, $\mu_h=0.493$, $\mu=0.007$, and $\delta=0.001$]{Temporal evolution of the soliton potential for $\kappa_c=7.0$, $\kappa_h=10.0$, $T_{ec}=10~eV$, $T_{eh}=100~eV$, $\mu_c=0.5$, $\mu_h=0.493$, $\mu=0.007$, and $\delta=0.001$.}
	\label{fig:soliton pot}
\end{figure}

In figure \ref{fig:soliton pot}, the solution of the solitary wave, reiterates the fact of soliton structure as it propagates throughout the medium mantaining its shape and size. However, the impact on individual species, sometimes is overshadowed by the potential. Since, the potential contains the information of density as well as the velocity of each species, it is not appropriate to come to a conclusion about the nature of the solitary wave by observing an undisturbed or perfectly propagating potential profile. Solving the same for ion density, the propagation of initial perturbation is also observed to be unaffected.

%\begin{figure}
%	\centering    
%	\includegraphics[width=0.8\textwidth]{figures/ion_den}
%	\caption[Propagation of the ion density for $\kappa_c=7.0$, $\kappa_h=10.0$, $T_{ec}=10~eV$, $T_{eh}=100~eV$, $\mu_c=0.5$, $\mu_h=0.493$, $\mu=0.007$, and $\delta=0.001$.]{Propagation of the ion density for $\kappa_c=7.0$, $\kappa_h=10.0$, $T_{ec}=10~eV$, $T_{eh}=100~eV$, $\mu_c=0.5$, $\mu_h=0.493$, $\mu=0.007$, and $\delta=0.001$.}
%	\label{fig:soliton ni}
%\end{figure}

\begin{figure}
	\centering    
	\includegraphics[width=0.8\textwidth]{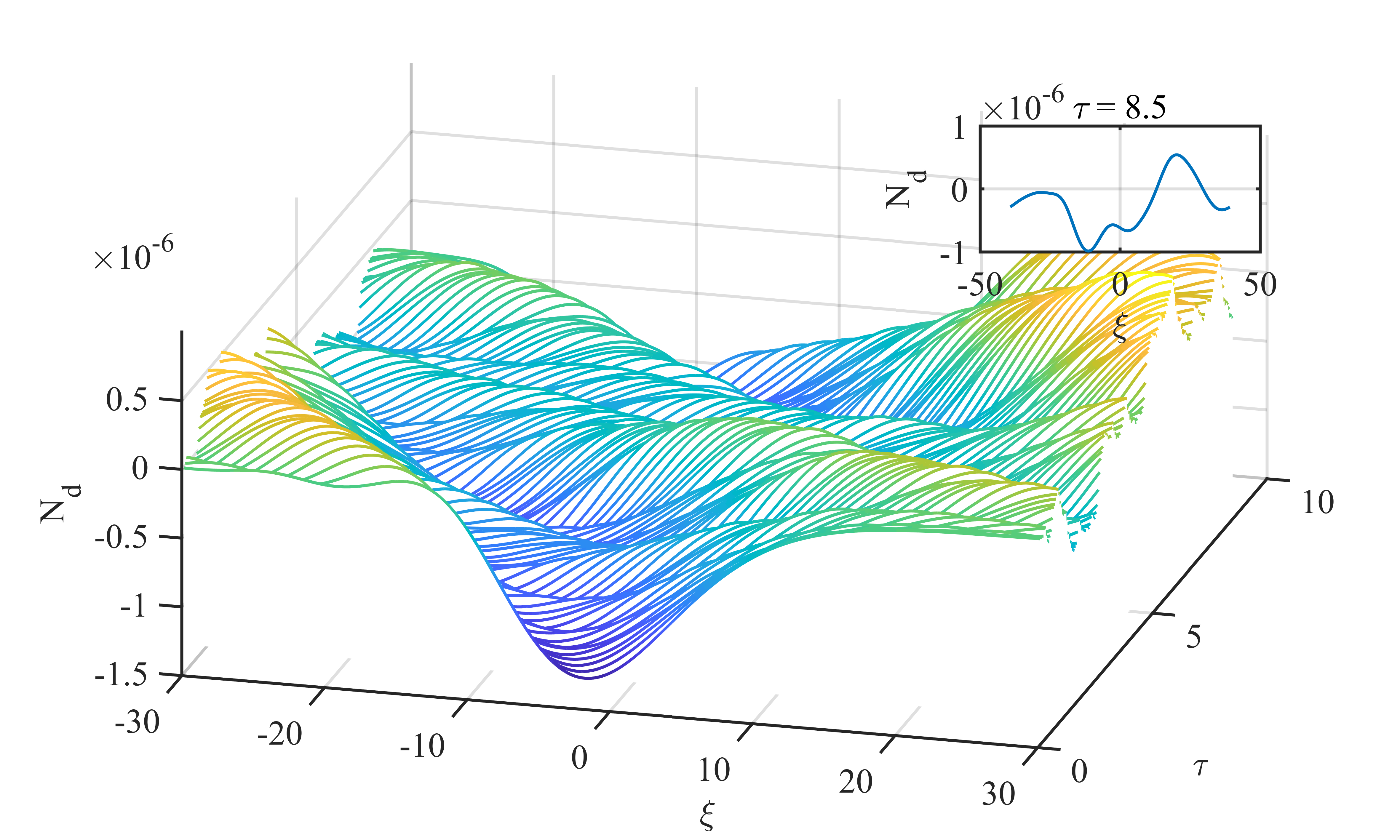}
	\caption[Propagation of the dust density for $\kappa_c=7.0$, $\kappa_h=10.0$, $T_{ec}=10~eV$, $T_{eh}=100~eV$, $\mu_c=0.5$, $\mu_h=0.493$, $\mu=0.007$, and $\delta=0.001$.]{Propagation of the dust density for $\kappa_c=7.0$, $\kappa_h=10.0$, $T_{ec}=10~eV$, $T_{eh}=100~eV$, $\mu_c=0.5$, $\mu_h=0.493$, $\mu=0.007$, and $\delta=0.001$.}
	\label{fig:soliton nd 1}
\end{figure}

\begin{figure}
	\centering    
	\includegraphics[width=0.8\textwidth]{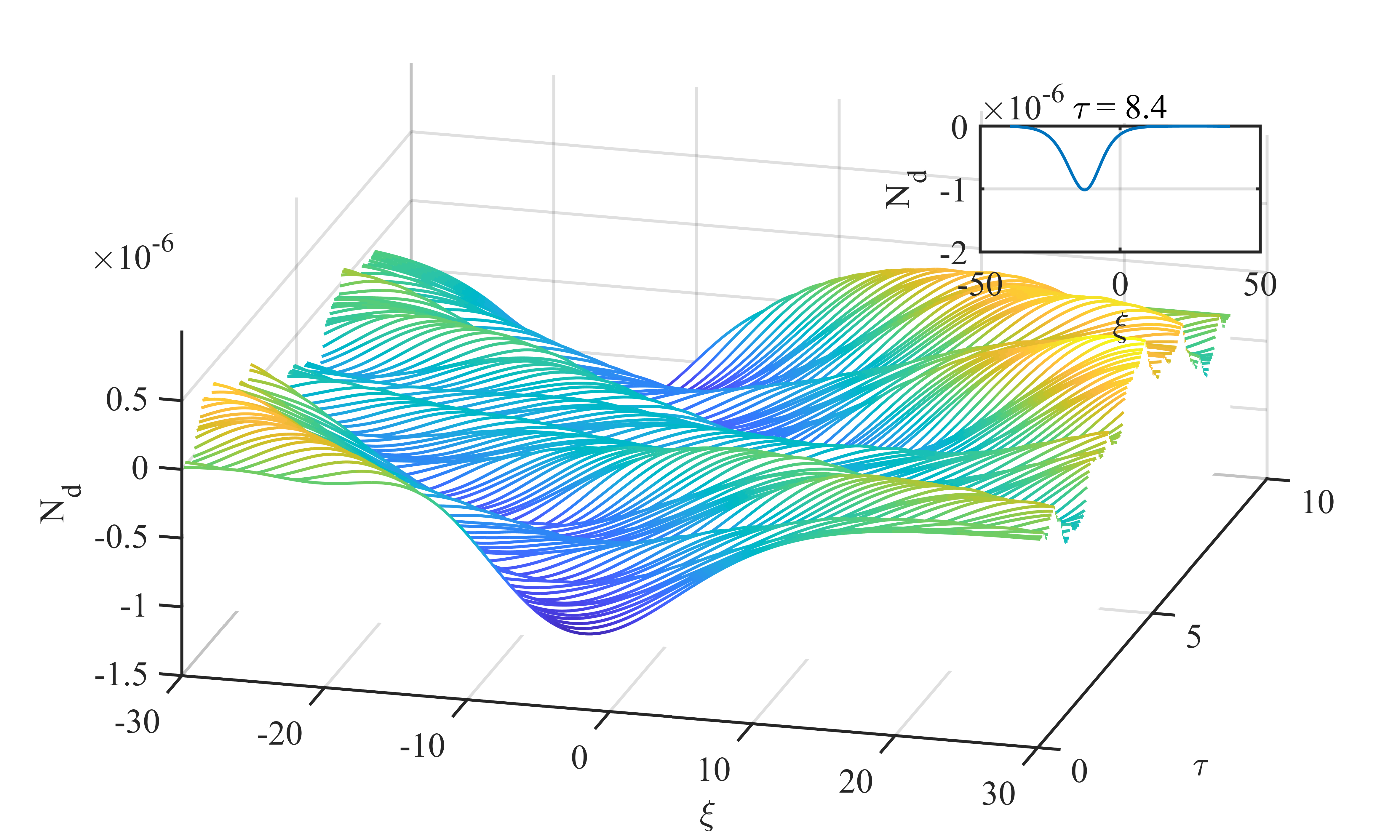}
	\caption[Propagation of the dust density for $\kappa_c=100.0$, $\kappa_h=100.0$, $T_{ec}=10~eV$, $T_{eh}=100~eV$, $\mu_c=0.5$, $\mu_h=0.493$, $\mu=0.007$, and $\delta=0.001$.]{Propagation of the dust density for $\kappa_c=100.0$, $\kappa_h=100.0$, $T_{ec}=10~eV$, $T_{eh}=100~eV$, $\mu_c=0.5$, $\mu_h=0.493$, $\mu=0.007$, and $\delta=0.001$.}
	\label{fig:soliton nd 2}
\end{figure}

\begin{figure}
	\centering    
	\includegraphics[width=0.8\textwidth]{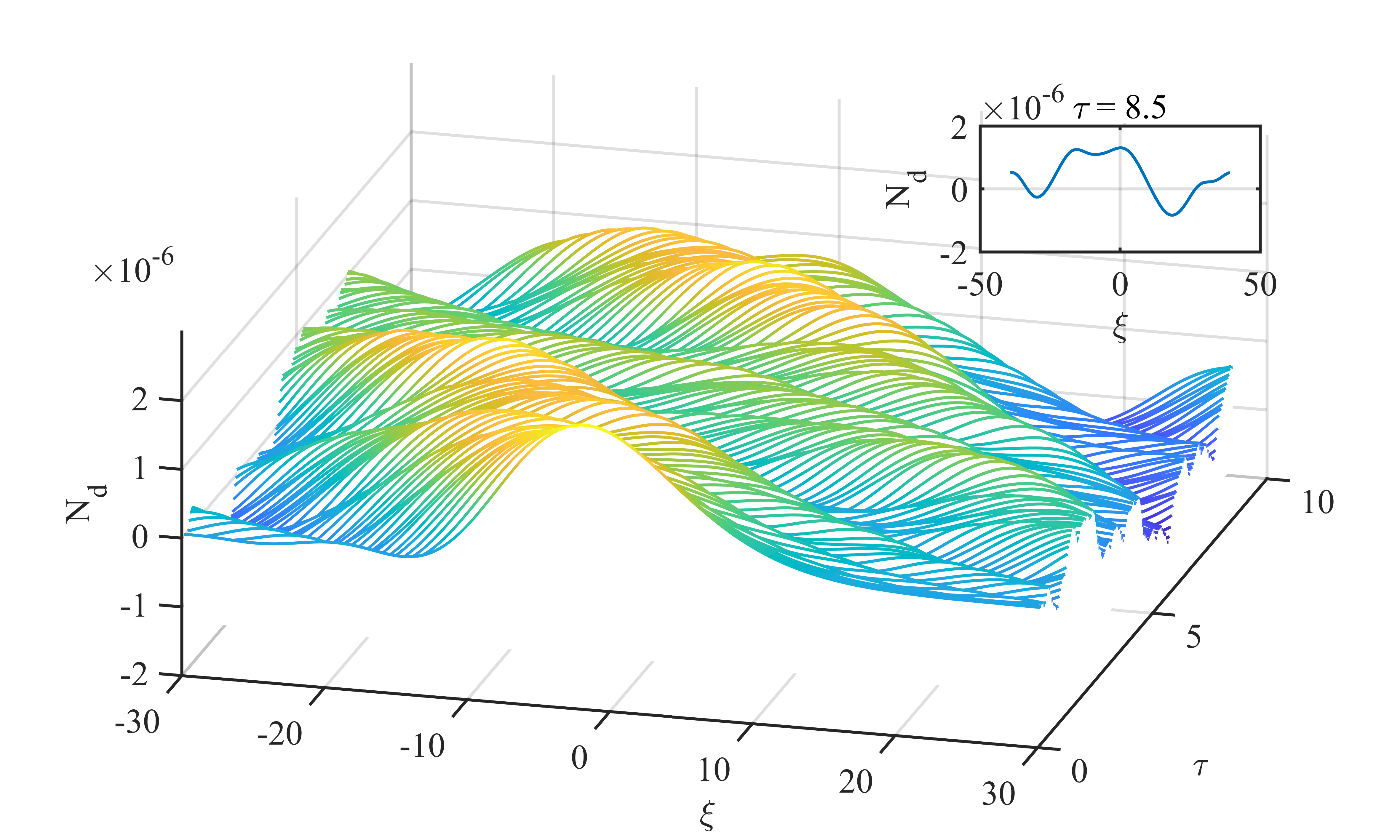}
	\caption[Propagation of the dust density for $\kappa_c=7.0$, $\kappa_h=10.0$, $T_{ec}=10~eV$, $T_{eh}=100~eV$, $\mu_c=0.1$, $\mu_h=0.893$, $\mu=0.007$, and $\delta=0.001$.]{Propagation of the dust density for $\kappa_c=7.0$, $\kappa_h=10.0$, $T_{ec}=10~eV$, $T_{eh}=100~eV$, $\mu_c=0.1$, $\mu_h=0.893$, $\mu=0.007$, and $\delta=0.001$.}
	\label{fig:soliton nd 3}
\end{figure}

\begin{figure}
	\centering    
	\includegraphics[width=0.8\textwidth]{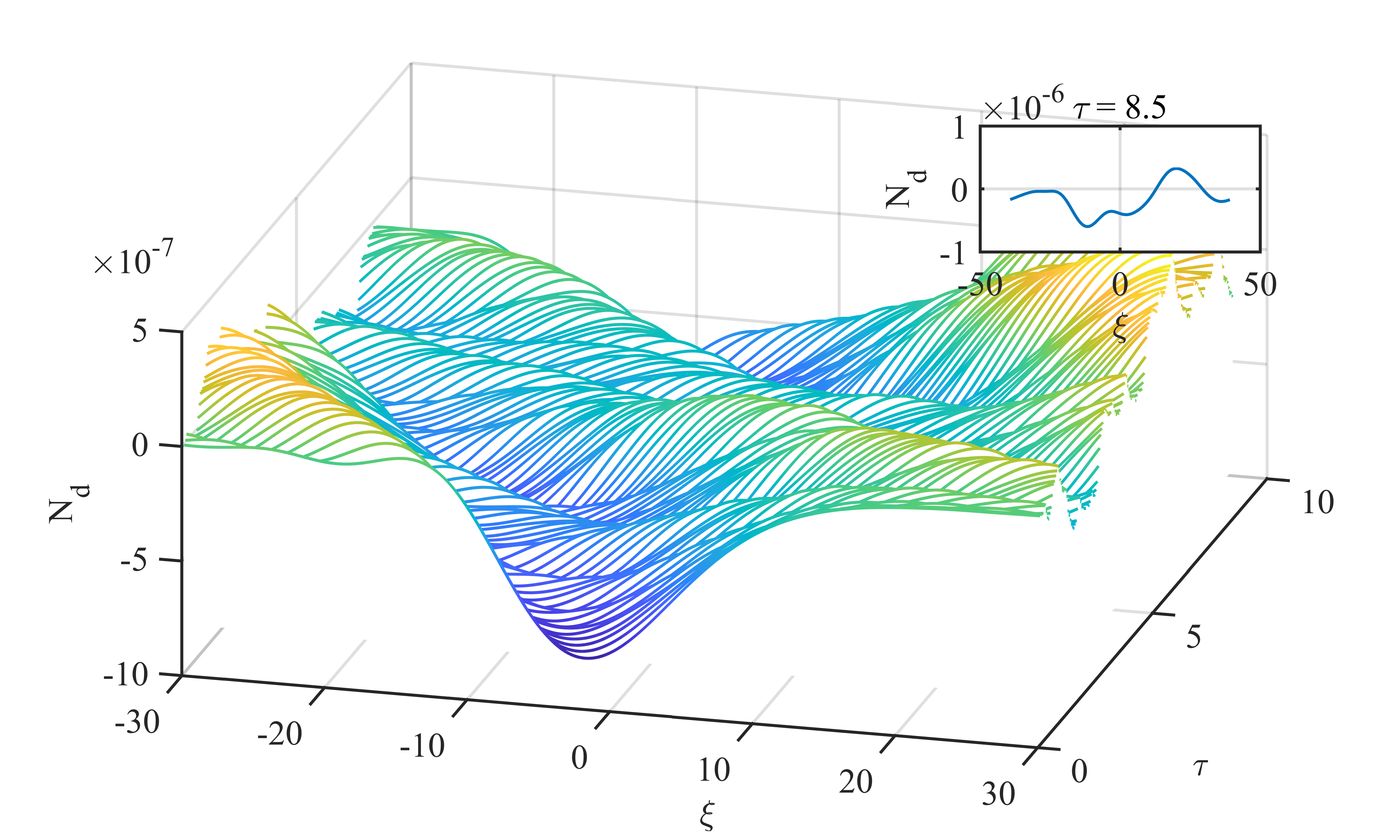}
	\caption[Propagation of the dust density for $\kappa_c=7.0$, $\kappa_h=10.0$, $T_{ec}=10~eV$, $T_{eh}=100~eV$, $\mu_c=0.9$, $\mu_h=0.093$, $\mu=0.007$, and $\delta=0.001$.]{Propagation of the dust density for $\kappa_c=7.0$, $\kappa_h=10.0$, $T_{ec}=10~eV$, $T_{eh}=100~eV$, $\mu_c=0.9$, $\mu_h=0.093$, $\mu=0.007$, and $\delta=0.001$.}
	\label{fig:soliton nd 4}
\end{figure}

\begin{figure}
	\centering    
	\includegraphics[width=0.8\textwidth]{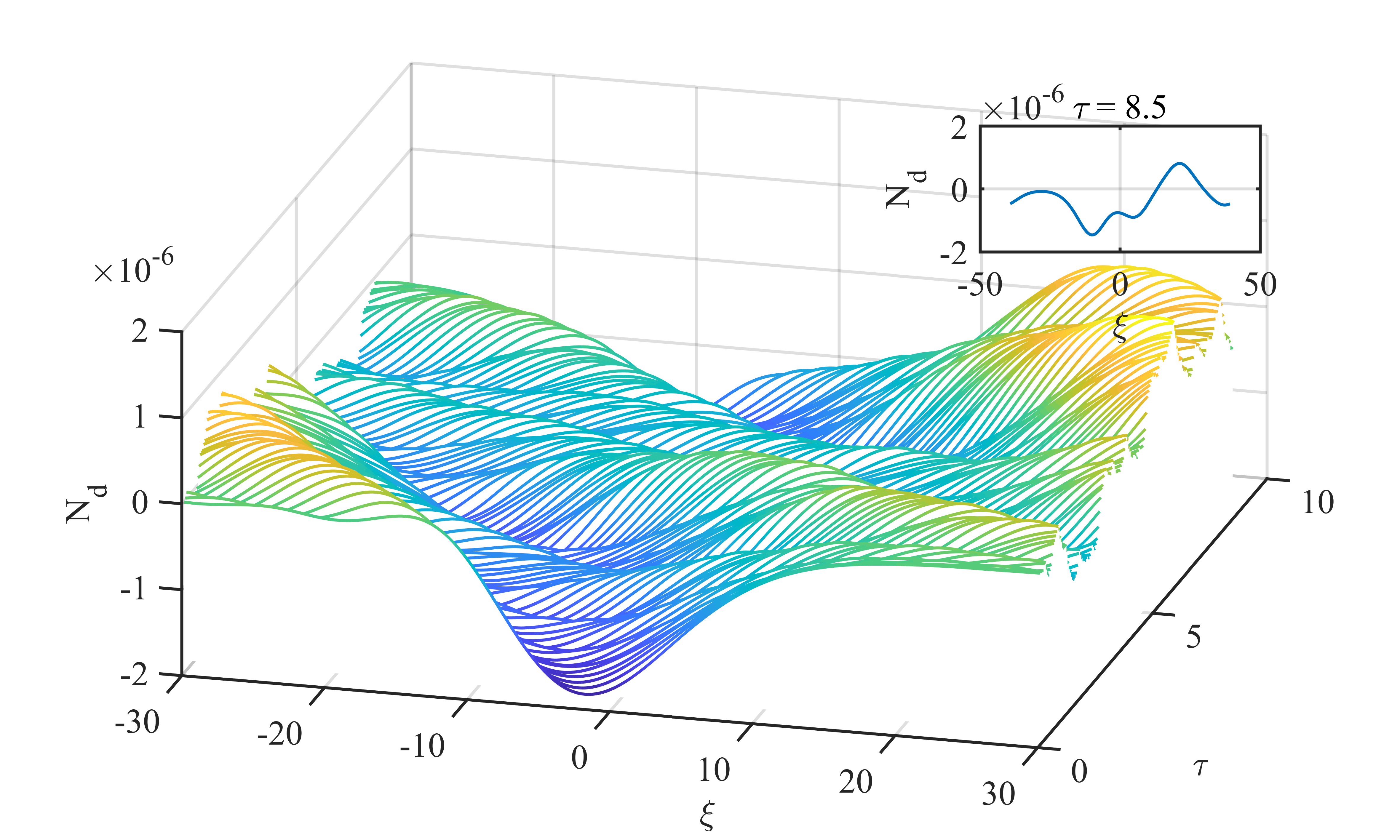}
	\caption[Propagation of the dust density for $\kappa_c=7.0$, $\kappa_h=10.0$, $T_{ec}=100~eV$, $T_{eh}=10~keV$, $\mu_c=0.5$, $\mu_h=0.493$, $\mu=0.007$, and $\delta=0.001$.]{Propagation of the dust density for $\kappa_c=7.0$, $\kappa_h=10.0$, $T_{ec}=100~eV$, $T_{eh}=10~keV$, $\mu_c=0.5$, $\mu_h=0.493$, $\mu=0.007$, and $\delta=0.001$.}
	\label{fig:soliton nd 5}
\end{figure}

\begin{figure}
	\centering    
	\includegraphics[width=0.8\textwidth]{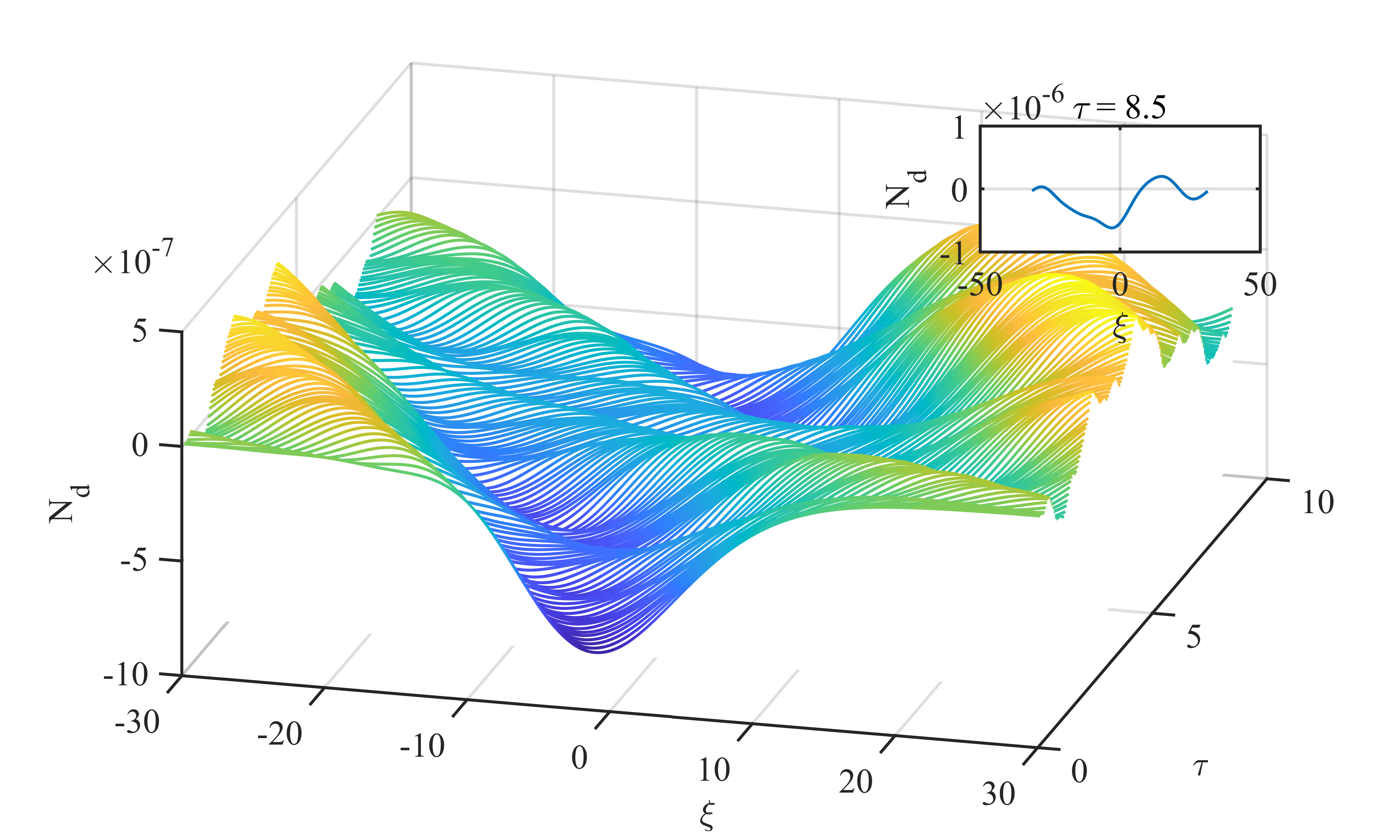}
	\caption[Propagation of the dust density for $\kappa_c=7.0$, $\kappa_h=10.0$, $T_{ec}=100~eV$, $T_{eh}=100~eV$, $\mu_c=0.5$, $\mu_h=0.493$, $\mu=0.007$, and $\delta=0.001$.]{Propagation of the dust density for $\kappa_c=7.0$, $\kappa_h=10.0$, $T_{ec}=100~eV$, $T_{eh}=100~eV$, $\mu_c=0.5$, $\mu_h=0.493$, $\mu=0.007$, and $\delta=0.001$.}
	\label{fig:soliton nd 6}
\end{figure}

Similarly, solving the KdV equation for the dust density with an initial density perturbation (equation (\ref{KdVd}))  has been shown in Figure \ref{fig:soliton nd 1}. Varying different parameters like spectral index, temperature ratio, and density ratio, it has been observed that the density profile cannot hold its shape and size while propagating, and is gradually damped. To extract the physics behind this anomalous behavior, the effects of different superthermal parameters on the dust density have been explored (figure \ref{fig:soliton nd 2} - \ref{fig:soliton nd 6}).

In figure \ref{fig:soliton nd 2} the evolution of dust density for Maxwellian electrons is depicted. The effective way of doing so is to consider the spectral indices as $\kappa_c=100$, and $\kappa_h=100$. For non-superthermal or Maxwellian electrons, it is perceived that, though the solitary wave initially splits up to multiple solitons, but with the passage of time, it regains its initial structure. It clearly indicates that the presence of superthermal particle in a plasma can drustically alter the nature of solitary waves in plasmas.

The density ratio of the superthermal electrons also may influence the nature of the proposed wave. In figure \ref{fig:soliton nd 3} and \ref{fig:soliton nd 4}, the space-time evolution of the dust density pertubation is observed by varying the concentration of the superthermal electrons. In the particular study, two different values of $\mu_c$ have been considered for two different cases. For the first case, $\mu_c=0.1$ (figure \ref{fig:soliton nd 3}), while for the second case, $\mu_c=0.9$ (figure \ref{fig:soliton nd 4}). The other  parameters are kept same as the initial study (figure \ref{fig:soliton nd 1}). From both the figures, it has been observed that the initial perturbation to the system gets disturbed. Hence, it can be inferred that the concentration of the electrons have played a significant role on the aberrant nature of the solitary wave. For higher concentration of the cold electron component, more number of peaks have been observed in the wave. Hence, the presence of large number of cold superthermal electrons in the plasma provides more disturbance to the dust density. Besides, it is also to be noted that there is a lower limit of $\mu_c$ for an undamped propagation of the potential. For a plasma with moderately superthermal electrons, the lower limit is found to be $0.3$. The value highly depends on the associated parameters like spectral index and temperature ratio. However, there is no upper bound for $\mu_c$. In fact, in the absence of the hot component of the electrons, the initial potential perturbation remains invarient with the passage of time.


The effect of two temperature electrons on the dust density is found quite interesting \cite{Schippers}. To put things in perspective, three different cases have been studied to analyze the space-time evolution of the dust density perturbation (figures \ref{fig:soliton nd 1}, \ref{fig:soliton nd 5}, \ref{fig:soliton nd 6}). In the first case, the minimum temperature for both of the electron species ($T_{ec}=10~eV$, $T_{eh}=100~eV$) have been considered (figure \ref{fig:soliton nd 1}). In the second case, the maximum temperatures for the electron species ($T_{ec}=100~eV$, $T_{eh}=10~keV$) have been considered (figure \ref{fig:soliton nd 5}). In the final case, the electrons have been assumed to be of equal temperature ($T_{ec}=T_{eh}=100~eV$) (figure \ref{fig:soliton nd 6}). The other  parameters are kept same as the initial study (figure \ref{fig:soliton nd 1}).
\begin{figure}
	\centering    
	\includegraphics[width=0.6\textwidth]{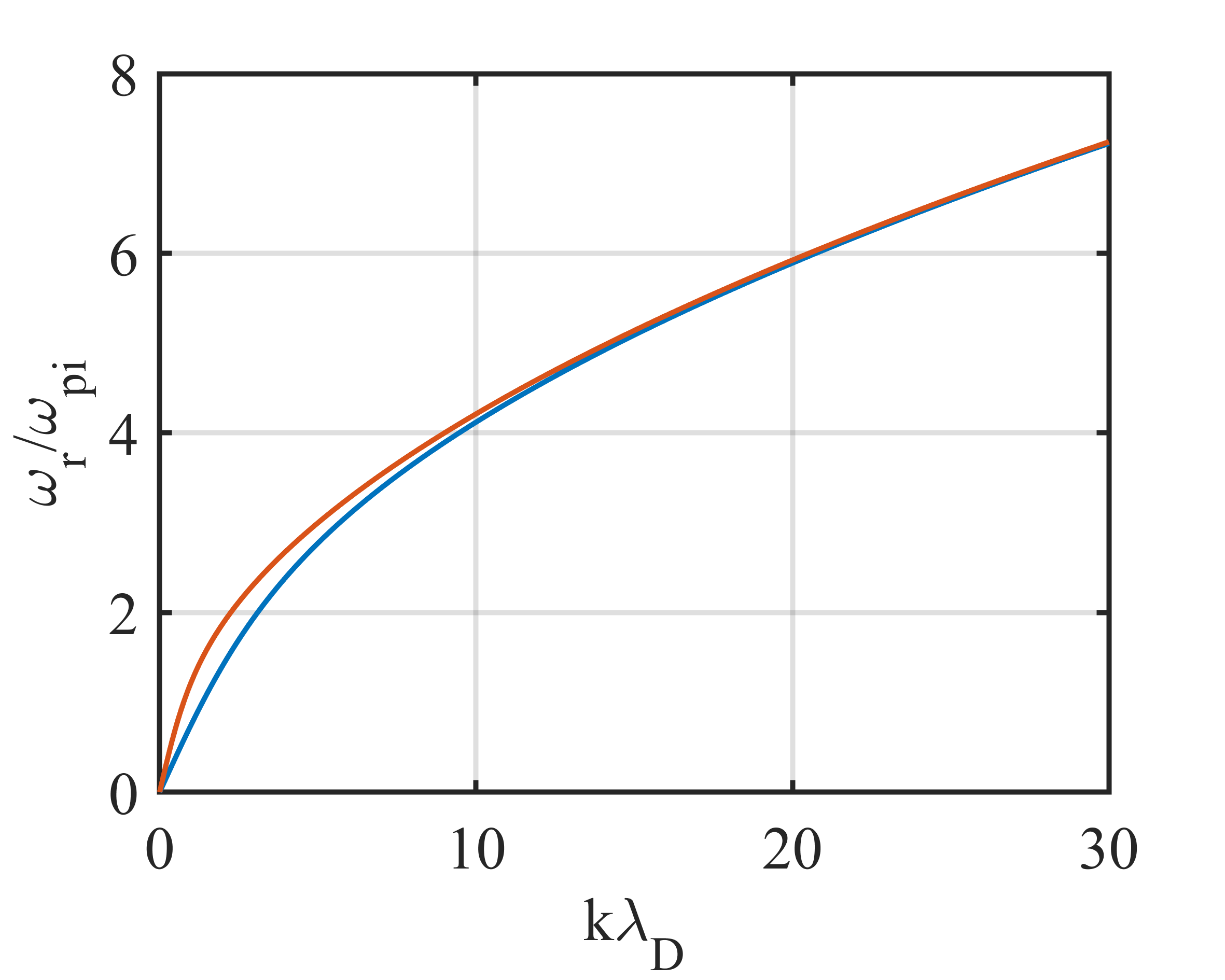}
	\caption{The dispersion curve of the DIA wave with $\mu_c=0.5$, $T_{ec}=10~eV$, $T_{eh}=100~eV$, $\delta=0.001$, $\mu=0.007$, and $(a)$ $\kappa_c=7.0$, $\kappa_h=10.0$ (blue line), $(b)$ $\kappa_c=100.0$, $\kappa_h=100.0$ (red line).}
	\label{fig:dispersion real}
\end{figure}
From the figures, it is quite clear that the density perturbation does not have the ablity to maintain its shape and size with the passage of time in all the three cases. However, for electrons with equal temperatures, it has been observed quite relaxed in comparison to other cases.

To justify the remarks obtained from the nonlinear analysis of the plasma waves, a linear analysis has been performed with the help of dispersion curves. The dispersion curve $k\lambda_D~-~\omega_r/\omega_{pi}$ is shown in figure \ref{fig:dispersion real} for two different sets of values of $\kappa_c$, and $\kappa_h$. From the previous studies, it has been observed that the spectral indices play a vital role in the damping of the wave. Hence, in this part of exploration, one component of the electrons have been considered as superthmal ($\kappa_c=7.0$, $\kappa_h=10.0$), while the other as Maxwellian ($\kappa_c=100$, $\kappa_h=100$). The other  parameters are kept same as the initial study (figure \ref{fig:soliton nd 1}). From the figure, it has been observed that the phase velocity of the wave increases with the increase in the values of the spectral indices of the electrons. This variation is more prominent in the short wavelength region. However, for $k\lambda_D\gg10$, the phase velocities of the waves become inseparable.
\begin{figure}
	\centering    
	\includegraphics[width=0.6\textwidth]{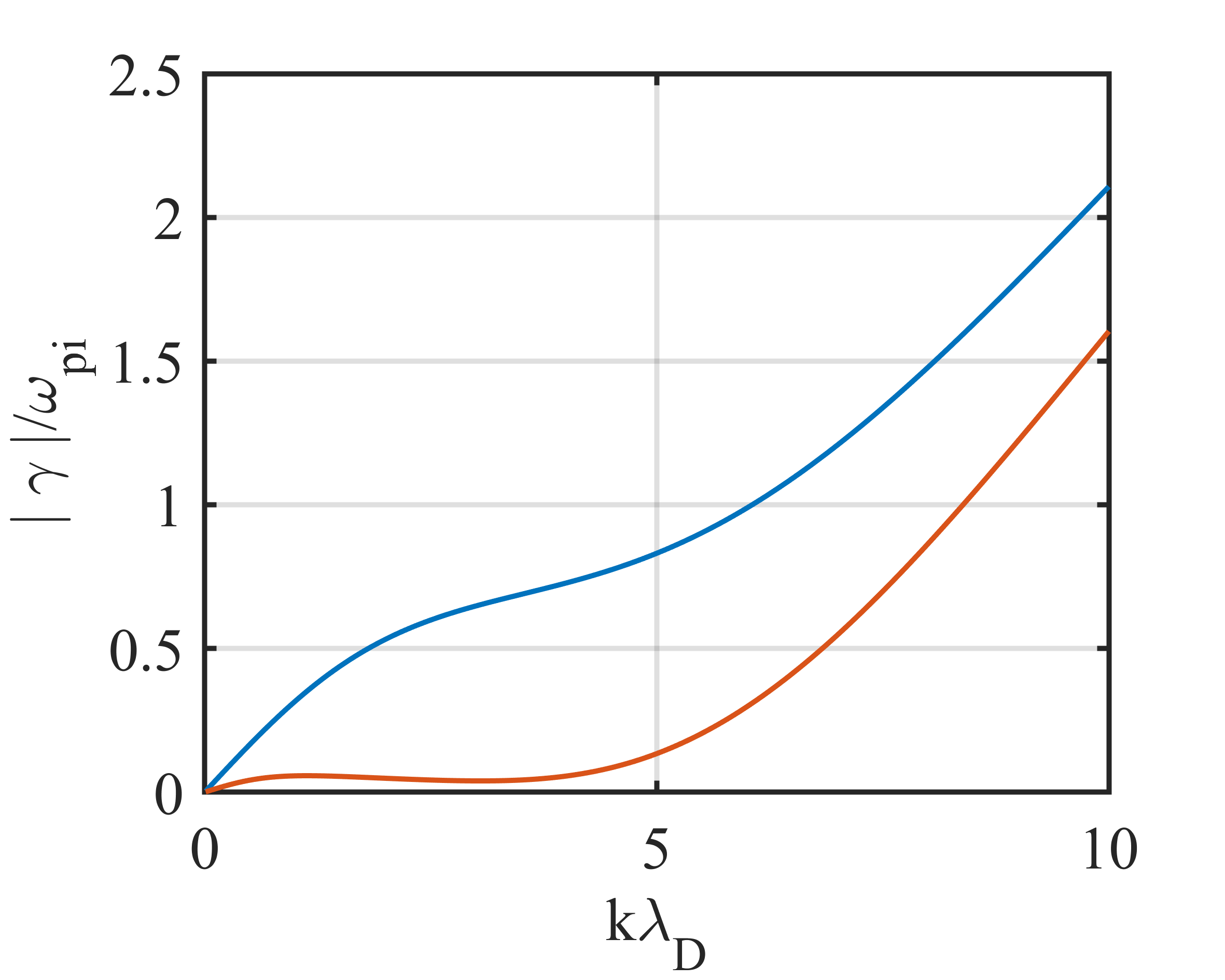}
	\caption{The wave number dispersion of the damping of the DIA wave with  $\mu_c=0.5$, $T_{ec}=10~eV$, $T_{eh}=100~eV$, $\delta=0.001$, $\mu=0.007$, and $(a)$ $\kappa_c=7.0$, $\kappa_h=10.0$ (blue line), $(b)$ $\kappa_c=100.0$, $\kappa_h=100.0$ (red line).}
	\label{fig:dispersion imag}
\end{figure}

The damping or growth in the wave can be proved by solving the linear dispersion relation (equation (\ref{dispersion 3 2})). In figure \ref{fig:dispersion imag}, the damping of the wave is plotted against the wave number by assuming the same set of spectral indices as mentined earlier. The figure reflects a comparison between the damping of the plasma waves with higher superthermal electrons (blue line), and with Maxwellian (red line) electrons. The damping in the first one is found to be greater than the sencond one. Remarkably, this behavior is analogous to the conclusions extracted from figure \ref{fig:soliton nd 1} – figure \ref{fig:soliton nd 2}. Thus, the outcomes are firm enough to justify the aforementioned studies and their conclusions.

The phase difference between the dust charge variation and the plasma wave is supposed to lead to a strong damping of the wave \cite{Melands}. Therefore, to check whether the dust charge variation is responsible for this aberrant nature of the proposed DIA wave, the KdV equation for dust density with fixed dust charge is derived. By solving the equation, it has been perceived that the initial perturbation to the dust density is disturbed similar to the ealier cases. Thus, it can be concluded that the self-consistent dust charging is not wholely responsible for the damping of the dust density in a plasma containing superthermal electrons. However, considering dust charge as a variable makes the model more realistic compared to the fixed one.

%\begin{figure}
%	\centering    
%	\includegraphics[width=0.7\textwidth]{figures/dust_den_fixed}
%	\caption[Propagation of the dust density for $\kappa_c=7.0$, $\kappa_h=10.0$, $T_{ec}=100~eV$, $T_{eh}=10~keV$, $\mu_c=0.5$, $\mu_h=0.493$, $\mu=0.007$, $\delta=0.001$ for fixed dust charge ($Z_{d0}=10^3$).]{Propagation of the dust density for $\kappa_c=7.0$, $\kappa_h=10.0$, $T_{ec}=100~eV$, $T_{eh}=10~keV$, $\mu_c=0.5$, $\mu_h=0.493$, $\mu=0.007$, $\delta=0.001$ for fixed dust charge ($Z_{d0}=10^3$).}
%	\label{fig:soliton fixed zd}
%\end{figure}

%********************************** %Fifth Section  **************************************
\section{Conclusions} \label{Section_5}

The influence of superthermal electrons on the propagation of Dust Ion Acoustic solitons is studied by numerically solving the KdV equation. The steady state solution of the KdV equation is used as an initial perturbation in the time dependent study. It has been observed that both the potential and ion number density can sustain the shape and size of the initial perturbation, while the dust density behaves differently.

The propagation of the solitary wave of the dust density is examined for different values of the superthermal parameters. It has been observed that the presence of superthermal electrons accelerates the damping of the wave. The dust density is also sensitive to the cold electron concentration in the plasma. More the number of cold electron component, more is the disturbance in the initial perturbation. Other than that, it also responds strongly to the temperature ratio of the electrons. Superthermal electrons with equal temperature results in a quite relaxed density profile. In summary, the presence of superthermal particles significantly alter the nature of the waves in the dusty plasmas.

The dispersion of the wave frequency and damping rate with the wave number seems to bring another perspective to the present theory. The presence of superthermal electrons appears to have larger role in the wave damping process.

Irregardless of the assumption of Dust Ion Acoustic plasma mode,  the observations are found to follow Dust Acoustic mode. Since ions are not responding to the superthemal parameters, it is believed, one can have the same results with the DA assumption instead of DIA.

The results obtained from the present investigation would be helpful in understanding the nonlinear structures in space dusty plasma like Saturn magnetosphere.

%*****************************************************************************************	


\section*{References}

\begin{thebibliography}{50}
	\bibitem{Vasyliunas} Vasyliunas V M 1968 Low-energy electrons on the day side of the magnetosphere \textit{J. Geophys. Res.} \textbf{73} 7519–23.
	
	\bibitem{Montgomery} Montgomery M D, Bame S J and Hundhausen A J 1968 Solar wind electrons: Vela 4 measurements \textit{J. Geophys. Res.} \textbf{73} 4999–5003.
	
	\bibitem{Feldman} Feldman W C, Asbridge J R, Bame S J, Montgomery M D and Gary S P 1975 Solar wind electrons \textit{J. Geophys. Res.} \textbf{80} 4181–96.
	
	\bibitem{Pierrard} Pierrard V and Lazar M 2010 Kappa Distributions: Theory and Applications in Space Plasmas \textit{Sol. Phys.} \textbf{267} 153–74.
	
	\bibitem{Lazar_Schli} Lazar M, Schlickeiser R and Poedts S 2012 \textit{Suprathermal Particle Populations in the Solar Wind and Corona Exploring the Solar Wind} (InTech) p 241.
	
	\bibitem{Maksimovic} Maksimovic M, Pierrard V and Lemaire J F 1997 A kinetic model of the solar wind with Kappa distribution functions in the corona \textit{Astron. Astrophys} \textbf{324} 725–34.
	
	\bibitem{Lazar} Lazar M, Kourakis I, Poedts S and Fichtner H 2018 On the effects of suprathermal populations in dusty plasmas: The case of dust-ion-acoustic waves \textit{Planet. Space Sci.} \textbf{156} 130–8.
	
	\bibitem{Alam} Alam M S, Masud M M and Mamun A A 2014 Effects of two-temperature superthermal electrons on dust-ion-acoustic solitary waves and double layers in dusty plasmas \textit{Astrophys. Space Sci.} \textbf{349} 245–53.
	
	\bibitem{Schippers} Schippers P, Blanc M, André N, Dandouras I, Lewis G R, Gilbert L K, Persoon A M, Krupp N, Gurnett D A, Coates A J, Krimigis S M, Young D T and Dougherty M K 2008 Multi-instrument analysis of electron populations in Saturn’s magnetosphere \textit{J. Geophys. Res. Sp. Phys.} \textbf{113} 1–10.
	
	\bibitem{Chunshi} Chunshi Cui and Goree J 1994 Fluctuations of the charge on a dust grain in a plasma \textit{IEEE Trans. Plasma Sci.} \textbf{22} 151–8.
	
	\bibitem{Baluku} Baluku T K and Hellberg M A 2015 Kinetic theory of dust ion acoustic waves in a kappa-distributed plasma \textit{Phys. Plasmas} \textbf{22} 083701.
	
	\bibitem{Moulick} Moulick R and Goswami K S 2014 Sheath formation under collisional conditions in presence of dust \textit{Phys. Plasmas} \textbf{21} 083702.
	
	\bibitem{Kakati} Kakati M and Goswami K S 1998 Solitary wave structures in presence of nonisothermal ions in a dusty plasma \textit{Phys. Plasmas} \textbf{5} 4508–10
	
	\bibitem{Dutta} Dutta D and Goswami K S 2019 Dust ion acoustic double layer in the presence of superthermal electrons \textit{Indian J. Phys.} \textbf{93} 257–65.
	
	\bibitem{Mishra1} Mishra R and Dey M 2018 Propagation of high frequency electrostatic surface waves along the planar interface between plasma and dusty plasma \textit{Phys. Scr.} \textbf{93} 045601.
	
	\bibitem{Shukla_mamun} Shukla P K and Mamun A A 2001 Dust-acoustic shocks in a strongly coupled dusty plasma \textit{IEEE Trans. Plasma Sci.} \textbf{29} 221–5.
	
	\bibitem{Mishra2} Mishra R and Dey M 2018 Propagation of electrostatic surface wave along the dust void boundary \textit{Phys. Scr.} \textbf{93} 085601.
	
	\bibitem{Shukla_Silin} Shukla P K and Silin V P 1992 Dust ion-acoustic wave \textit{Phys. Scr.} \textbf{45} 508–508.
	
	\bibitem{Nejoh} Nejoh Y N 1997 The dust charging effect on electrostatic ion waves in a dusty plasma with trapped electrons \textit{Phys. Plasmas} \textbf{4} 2813–9.
	
	\bibitem{Melands} Melandsø F, Aslaksen T and Havnes O 1993 A new damping effect for the dust-acoustic wave \textit{Planet. Space Sci.} \textbf{41} 321–5.
	
	\bibitem{Swanson} Swanson D G 2003 \textit{Series in Plasma Physics Plasma Waves} (Institute of Physics Publishing, Bristol and Philadelphia) p 366.
	
	\bibitem{Masood} Masood W, Mushtaq A and Khan R 2007 Linear and nonlinear dust ion acoustic waves using the two-fluid quantum hydrodynamic model \textit{Phys. Plasmas} \textbf{14} 123702.
	
	\bibitem{Xie} Xie B, He K and Huang Z 1999 Dust-acoustic solitary waves and double layers in dusty plasma with variable dust charge and two-temperature ions \textit{Phys. Plasmas} \textbf{6} 3808–16
	
	\bibitem{Shukla_book} Shukla P K and Mamun A A 2002 \textit{Introduction to Dusty Plasma Physics} (Institute of Physics Publishing, Bristol and Philadelphia) p 37.
	
	\bibitem{Hakimi} Hakimi Pajouh H and Afshari N 2016 Influence of superthermal plasma particles on the Jeans instability in self-gravitating dusty plasmas with dust charge variations \textit{Phys. Lett. A} \textbf{380} 3810–6.
	
	\bibitem{Bora} Bora M P, Choudhury B and Das G C 2012 Dust-acoustic soliton in a plasma with multi-species ions and kappa-distributed electrons \textit{Astrophys. Space Sci.} \textbf{341} 515–25.
	
	\bibitem{Duha} Duha S S and Mamun A A 2009 Dust-ion-acoustic shock waves due to dust charge fluctuation \textit{Phys. Lett. A} \textbf{373} 1287–9.
	
	\bibitem{Abramowitz} Abramowitz M and Stegun I A 1972 \textit{Handbook of Mathematical Functions with Formulas, Graphs, and Mathematical Tables} (National Bureau of Standards, Washington) p 556.
	
	\bibitem{Fried} Fried B D and Conte S D 1961 \textit{The Plasma Dispersion Function} (Academic Press, New York) p 1.
	
	\bibitem{Chen} Chen F F 1983 \textit{Introduction to Plasma Physics and Controlled Fusion} (Plenum Press, Newyork and London) p 211.
	
	\bibitem{Kassam} Kassam A-K and Trefethen L N 2005 Fourth-Order Time-Stepping for Stiff PDEs \textit{SIAM J. Sci. Comput.} \textbf{26} 1214–33.
	
	\bibitem{Mishra_Corr} Mishra R, Adhikari S, Mukherjee R and Dey M 2018 Correlation between two non-linear events in a dusty plasma system \textit{Phys. Plasmas} \textbf{25} 123703.
	
	
	
	
	
	
	
	
	
	
	
	
	
	
	
	
	
	
	
	
	
	
	

\end{thebibliography}
\end{document}